\documentclass[twocolumn,aps,pra,groupedaddress,superscriptaddress,amsmath,amssymb,noeprint,nofootinbib]{revtex4-2}
\usepackage[T1]{fontenc}
\usepackage{graphicx}
\graphicspath{{./Figures/}}
\usepackage{dcolumn}
\usepackage{bm}
\usepackage[dvipsnames]{xcolor}
\usepackage{hyperref}
\hypersetup{colorlinks,linkcolor={MidnightBlue},citecolor={MidnightBlue},urlcolor={MidnightBlue}}  

\usepackage{float}

\usepackage{physics}
\usepackage{amsfonts}
\usepackage{verbatim}
\usepackage{amsmath}
\usepackage{csquotes}
\usepackage{color}
\usepackage{soul}
\usepackage{amsthm}
\usepackage{bm}
\usepackage{graphicx}
\usepackage[normalem]{ulem}

\usepackage{verbatim}

\newenvironment{equations}
{\begin{equation}\begin{aligned}}
		{\end{aligned}\end{equation}\ignorespacesafterend}

\newcommand{\mq}[2]{\abs{\ip{#1}{#2}}^2}
\newcommand{\prt}[1]{\left(#1\right)}

\newcommand{\intmp}{\int_{-\infty}^{+\infty}}

\newcommand{\Schr}{Schr\"{o}dinger\ }

\newcommand{\dg}{\dagger}

\newcommand{\prtq}[1]{\left[#1\right]}

\newcommand{\prtg}[1]{\left\{#1\right\}}

\newcommand{\prtqB}[1]{\Bigg[#1\Bigg]}

\newcommand{\prtgB}[1]{\Bigg\{#1\Bigg\}}

\newcommand{\mcE}{\mathcal{E}}

\newcommand{\mcT}{\mathcal{T}}

\newcommand{\mcS}{\mathcal{S}}

\newcommand{\mcI}{\mathcal{I}}

\newcommand{\sigt}{\sigma_\Theta}

\newcommand{\tpsi}{\tilde{\psi}}

\newcommand{\Id}{\mathbb{I}}

\newcommand{\ad}{a^\dagger}

\newcommand{\cd}{c^\dagger}

\newcommand{\hn}{\hat{n}}

\newcommand{\tH}{\tilde{H}}

\newcommand{\tU}{\tilde{U}}
\newcommand{\tV}{\tilde{V}}

\newcommand{\nAv}{\bar{n}}

\begin{document}

\title{Fundamental mechanisms of energy exchanges in autonomous measurements based on dispersive qubit-light interaction}

\author{Nicol\`{o} Piccione}
\email{nicolo'.piccione@units.it}
\affiliation{Department of Physics, University of Trieste, Strada Costiera 11, 34151 Trieste, Italy}
\affiliation{Istituto Nazionale di Fisica Nucleare, Trieste Section, Via Valerio 2, 34127 Trieste, Italy}

\author{Maria Maffei}
\affiliation{Dipartimento di Fisica, Università di Bari, I-70126 Bari, Italy}

\author{Xiayu Linpeng}
\affiliation{International Quantum Academy, Futian District, Shenzhen 518048, China}
\affiliation{Department of Physics, Washington University, St. Louis, Missouri 63130, USA}

\author{Andrew N. Jordan}
\affiliation{The Kennedy Chair in Physics, Chapman University, Orange, CA 92866, USA}
\affiliation{Institute for Quantum Studies, Chapman University, Orange, California 92866, USA}
\affiliation{Department of Physics and Astronomy, University of Rochester, Rochester, New York 14627, USA}

\author{Kater W. Murch}
\affiliation{Department of Physics, Washington University, St. Louis, Missouri 63130, USA}

\author{Alexia Auff\`{e}ves}
\affiliation{Université Grenoble Alpes, CNRS, Grenoble INP, Institut Néel, 38000 Grenoble, France}
\affiliation{MajuLab, CNRS–UCA-SU-NUS-NTU International Joint Research Laboratory}
\affiliation{Centre for Quantum Technologies, National University of Singapore, 117543 Singapore, Singapore}

\begin{abstract}

Measuring an observable which does not commute with the Hamiltonian of a quantum system usually modifies the mean energy of this system. In an autonomous measurement scheme, coupling the system to a quantum meter, the system's energy change must be compensated by the meter's energy change. Here, we theoretically study such an autonomous meter-system dynamics: a qubit interacting dispersively with a light pulse propagating in a one-dimensional waveguide. The phase of the light pulse is shifted, conditioned to the qubit's state along the $z$-direction, while the orientation of the qubit Hamiltonian is arbitrary.
As the interaction is dispersive, photon number is conserved so that energy balance has to be attained by spectral deformations of the light pulse. Building on analytical and numerical solutions, we reveal the mechanism underlying this spectral deformation and display how it compensates for the qubit's energy change. We explain the formation of a three-peak structure of the output spectrum and we provide the conditions under which this is observable.
\end{abstract}

\maketitle

\section{Introduction}

In quantum mechanics, measurements of observables which do not commute with the system's bare Hamiltonian are ubiquitous, a paradigmatic example being the measurement of a free particle's position.
Recently, such measurements have been investigated as fuel for quantum engines, in which the measurement channels are analogous to hot thermal baths~\cite{Elouard2017Extracting,Yi2017SingleTemperature,Mohammady2017Szilard,Elouard2018Efficient,Ding2018MeasurementDriven,Buffoni2019MeasurementCooling,Jordan2020MeasurementEnginesInterpretation,Bresque2021Two-Qubit,Stevens2022Energetics,Yanik2022ThermodynamicsMeasurement,Jussiau2023ManyBody}. In these setups, the system's energy change due to the measurement has been dubbed \textit{quantum heat}~\cite{Elouard2017Role} or \textit{measurement energy}~\cite{Rogers2022Postselection}.
Following von Neumann's description~\cite{jordan2024quantum}, a quantum measurement primarily arises from a \textit{pre-measurement}, i.e., a closed-system interaction between the measured system and a second one called a \textit{quantum meter}. Further insights about the energy transfers happening during a measurement can be gained by studying the pre-measurement dynamics~\cite{Bresque2021Two-Qubit,Linpeng2023QuantumEnergetics}. When the pre-measurement features a scattering-type dynamics~\cite{Book_Taylor2006Scattering}, i.e., the system-meter interaction Hamiltonian is time-independent and autonomously turns on and off, the sum of the system's and meter's bare energies is the same at the beginning and at the end of the process. Hence, the measurement energy is conserved and it is balanced by the meter's energy change.

A perfect platform to analyze scattering-type pre-measurements in quantum optics is given by an electromagnetic pulse that interacts dispersively with a qubit~\cite{Blais2004Architecture,Book_Haroche2006,Blais2007Processing,Schuster2007resolving,Gambetta2008Zeno,Boissonneault2009Dispersive,Blais2021CQED,jordan2024quantum}. In this setting, the electromagnetic field features the quantum meter and the dispersive interaction induces a phase-shift of the light pulse, conditioned to the qubit's state along the $z$-axis: this provides a pre-measurement of the latter. 
Moreover, the pulse propagation allows one to autonomously turn on and off the coupling between the light and the qubit.
In the standard dispersive readout, the qubit's bare Hamiltonian is proportional to $\sigma_z$ and hence commutes with both the interaction Hamiltonian and the measured observable. This kind of dispersive measurements is ubiquitous in circuit QED~\cite{Blais2021CQED,jordan2024quantum}.

Here we consider this dispersive pre-measurement scenario when the qubit's bare Hamiltonian is tilted with respect to the $z$-axis by an arbitrary angle $\Theta$. This implies that the measured observable ($\sigma_z$) and the qubit's bare Hamiltonian do not commute, giving rise to an energy exchange between qubit and meter.
Remarkably, the dispersive interaction forbids a direct energy exchange via creation and/or annihilation of photons. Hence, the field's energy change manifests through spectral deformations, red shifts and blue shifts, while the photon number remains the same.

We analytically solve the joint qubit-field dynamics and use our solution to characterize the mechanism underlying the spectral deformation of the scattered pulse. We find conditions under which the deformation leads to a spectrum comprising three peaks whose positions are determined by the qubit's bare frequency and the pulse's carrier frequency. Importantly, the peaks do not depend on the intensity of the light pulse, as opposed to Mollow triplets~\cite{Book_Breuer2002,Book_Walls2007}. 
While Mollow triplets arise from the direct exchange of excitations between the qubit and the field, the spectral deformation under study stems from the dispersive interaction and is due to the non-commutation condition.
The experimental observation of a three-peak spectrum in the non-commuting dispersive regime has been recently reported in Ref.~\cite{Linpeng2023QuantumEnergetics}. 

Eventually, using Gaussian coherent pulses, we compute output spectra in various configurations, verifying the accuracy of our predictions. The pulses used to plot the spectra have long duration compared to the qubit's typical timescale. Indeed, short pulses lead to accurate measurements in the $\sigma_z$-basis but, simultaneously, the spectrum of the scattered field does not show any visible change; on the contrary, according to the WAY theorem~\cite{Loveridge2011Measurement} (named after Wigner, Araki, and Yanase), long pulses lead to bad measurements but allow one to resolve the scattered pulse's spectral deformation providing access to the energy change induced by the pre-measurement (or measurement energy).

The paper is organized as follows. In Sec.~\ref{Sec:Model}, we present the model, display the working-mechanism of the dispersive readout, and provide the formal solution of the joint qubit-field dynamics. In Sec.~\ref{Sec:DynamicsSOlution}, the paper's core, we analyze the mechanism underlying the spectral deformation building on the formal solution derived in Sec.~\ref{Sec:Model}. We explain why the final spectrum exhibits a visible three-peak structure becoming more and more visible as the frequency dispersion of the incoming pulse becomes smaller and smaller with respect to a quantity determined by the qubit's bare frequency and the photon number. In Sec.~\ref{sec:CoherentPulses}, using Gaussian coherent pulses as quantum meters,
we numerically solve the pre-measurement dynamics and we show the spectral deformation visible for long pulses. Finally, in Sec.~\ref{Sec:Conclusions} we draw conclusions on our paper.

\section{System and model\label{Sec:Model}}

\subsection{Physical model}

\begin{figure}
\centering
\includegraphics[width=0.48\textwidth]{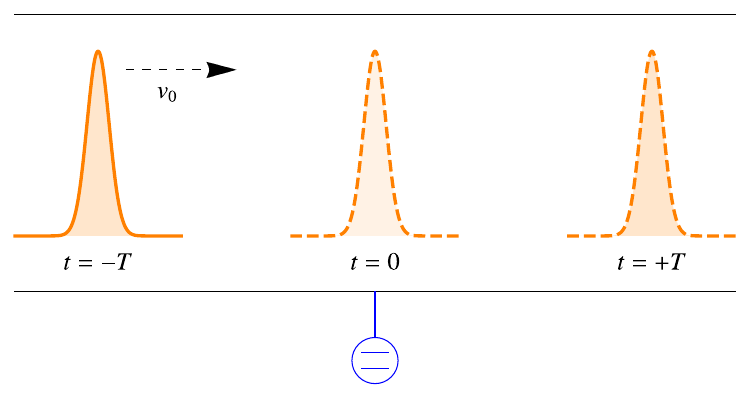}
\caption{Pictorial representation of the system under investigation. The traveling probe pulse is on the left of the qubit at $t=-T$ and on its right at $t=+T$. During this time, the pre-measurement takes place.}
\label{fig:FigWaveguide}
\end{figure}

Our system consists of a qubit dispersively coupled to an electromagnetic field in a one-dimensional waveguide (see Fig.~\ref{fig:FigWaveguide}). A probe pulse travels along the waveguide and its purpose is to measure the qubit in the $\sigma_z$-basis. Due to the waveguide's linear dispersion and chirality, the probe pulse travels with constant velocity $v_0$ from left to right. The pulse is on the left of the qubit at the initial time $t=-T$ and on its right at the final time $t=+T$. In this setting, the observable that we want to measure is the qubit's $\sigma_z$ and the electromagnetic field in the waveguide represents the quantum meter.

The Hamiltonian $H_Q$ of the qubit is given by
\begin{equation}
\label{eq:QubitBareHamiltonian}
H_Q = \frac{\hbar \omega_q}{2}\sigt,
\quad
\sigt = \cos(\Theta) \sigma_z + \sin(\Theta) \sigma_x,
\end{equation}
where $\omega_q$ is the qubit's frequency, $\sigma_x$ and $\sigma_z$ are Pauli matrices, and $\Theta$ quantifies how much the qubit's Hamiltonian is tilted with respect to $(\hbar\omega_q/2) \sigma_z$, which we refer to as the \enquote{non-tilted Hamiltonian}. The qubit is located at position $x=0$ inside the waveguide and is considered to be point-like.

The bare Hamiltonian of the electromagnetic field in the waveguide is a reservoir of electromagnetic modes of different frequencies~\cite{Gardiner1985}:
\begin{equation}
\label{eq:WaveguideHamiltonian}
H_{F} = \hbar \intmp \dd{k} \omega_k a^\dg_k a_k,
\qq{with}
\comm{a_k}{a^\dg_{k'}} = \delta (k - k'),
\end{equation}
where $\omega_k$ is the frequency of the mode associated with the wavevector $k$, $a_k$ and $\ad_k$ are respectively creation and annihilation operators of the bosonic mode with wavevector $k$, and $\delta(k)$ is the Dirac delta. Notice that we neglect the polarization degree of freedom of the propagating light. Assuming linear dispersion for all relevant field frequencies around the carrier frequency $\omega_0$, i.e., $\omega_k \simeq \omega_0 + v_0 k$, and exploiting the fact that the field propagates in one direction $(k>0)$, the above Hamiltonian can be cast into the following form [see Ref.~\cite{Shen2009Theory} and the Appendix~\ref{APPSec:FreeHamiltonian} for more details]:
\begin{equation}
\label{eq:WaveguideHamiltonianSpaceRepresentation}
H_F =  \intmp \dd{x} \prtq{\hbar \omega_0 \ad(x) a(x) - i \hbar v_0 \ad(x) \partial_x a(x)},
\end{equation}
where $a(x)$ and $\ad(x)$ are bosonic operators creating and destroying photons at position $x$ and satisfying the commutation relation $\comm{a(x)}{\ad(y)} = \delta (x-y)$. They are defined as follows:
\begin{equations}
\label{eq:BosonicOperatorsInverseTransformation}
a (x) &\equiv \frac{1}{\sqrt{2\pi}}\intmp \dd{k} a_k e^{+ikx},
\\
a^\dg (x) &\equiv \frac{1}{\sqrt{2\pi}}\intmp \dd{k} \ad_k e^{-ikx}.
\end{equations}
Finally, the dispersive interaction between qubit and field is given by
\begin{equation}
\label{eq:InteractionHamiltonian}
V = \intmp \dd{x} f(x) \ad(x) a(x) \otimes H_1,
\quad
H_1 \equiv \prtq{\frac{\hbar\chi_0}{2}}{\sigma_z},
\end{equation}
where $\chi_0$ represents the coupling strength between qubit and field while $f(x)$ represents the shape of the interaction region between them. Since the qubit is treated as point-like, from now on we will consider the case $f(x) \rightarrow L \delta (x)$, where $L$ is the effective length of the space occupied by the qubit.

We assume that the pulse is far on the left of the qubit at the initial time $t=-T$ so that they are in a product state at that time and their interaction takes place for $+T>t>-T$.
We write the initial state of the system in the following form:
\begin{multline}
\label{eq:FreeFieldNumberDecomposition}
\ket{\psi (-T)} = 
\prtqB{\beta_0 + \sum_{n=1}^{\infty} \frac{\beta_n}{\sqrt{n!}} \times \\ \times \intmp \dd{x^n} \phi_n (\vec{x}_n) A^\dg_n (\vec{x}_n)}\ket{0_F}\ket{\psi_Q (-T)},
\end{multline}
where $\ket{0_F}$ is the electromagnetic vacuum and $\ket{\psi_Q (0)}$ is the initial qubit's state. The coefficients $\beta_n$ are the amplitudes associated to each photon-number subspace, and satisfy $\sum_n \abs{\beta_n}^2 = 1$. In addition, we have defined $A^\dg_n (\vec{x}_n) \equiv \ad(x_1) \ad(x_2) \dots \ad(x_n)$, $\vec{x}_n \equiv x_1,x_2,\dots,x_n$, and $\dd{x^n} \equiv \dd{x_1} \dd{x_2} \dots \dd{x_n}$. The wavefunction $\phi_n (\vec{x}_n)$ is the probability density amplitude for the $n$-photon subspace, it is normalized (i.e, $\intmp \dd{x}^n \abs{\phi_n(\vec{x}_n)}^2 = 1$) and it is invariant for exchange of variables.

Let us notice that, due to the linear dispersion, we can equivalently express the operators $a(x)$ in terms of the time $t=-(T+x/v_0)$ at which a photon that was in the position $x < 0$ at the time $t=-T$ arrives at the qubit's position $x=0$. Hence, with a notation abuse, the operator $\ad(t)=\ad(x=-v_0 (T+t))$ creates a photon at time $t$ in the qubit's position $x=0$. Hereafter, we will use this time notation to simplify the interpretation of the dynamics. In this notation, the initial state reads
\begin{multline}
\label{eq:InitialState}
\ket{\psi(-T)} = \prtqB{\beta_0 + \sum_{n=1}^\infty \frac{\beta_n}{\sqrt{n!}}\times 
\\
\times \intmp \dd{t}^n \psi_n (\vec{t}_n)A^\dg_n (\vec{t}_n)}\ket{0_F}\ket{\psi_Q (-T)},
\end{multline}
where we have defined $A^\dg_n (\vec{t}_n) \equiv \ad(t_1),\ad(t_2),\dots,\ad(t_n)$, $\vec{t}_n \equiv t_1,t_2,\dots,t_n$, and $\dd{t}^n \equiv \dd{t_1} \dd{t_2} \dots \dd{t_n}$. Finally, the wavefunction $\psi_n (\vec{t}_n)$ is the analog of $\phi_n(\vec{x}_n)$.

\subsection{Dispersive read-out}

Let us display the measurement's mechanism in the simplest possible scenario where $\Theta=0$ and the field is prepared in a coherent state $\ket{\alpha}$. Let us consider the time $T$ at which the pulse has completely gone from the left to the right of the qubit. The dynamics then gives rise to the following map:
\begin{equations}
\label{eq:nontilt}
U(T,-T)\ket{e_{z}}\ket{\alpha}&= \ket{e_{z}}\ket{e^{-i\phi/2}\alpha};\\
U(T,-T)\ket{g_{z}}\ket{\alpha}&= \ket{g_{z}}\ket{e^{+i\phi/2}\alpha},
\end{equations}
where $U(T,-T)$ is the unitary operator evolving the system from $t=-T$ to $t=+T$,
$\ket{e_{z}}$ and $\ket{g_{z}}$ are eigenstates of $\sigma_z$, and the angle $\phi$ is given by $\phi=\chi_0 L/v_0$. Let us notice that the interaction leaves the eigenstates of $\sigma_z$ unchanged and therefore, the measurement process is repeatable because the observable's eigenstates are left unvaried by the pre-measurement dynamics. This is a quantum non-demolition (QND) measurement~\cite{Book_Haroche2006}. After the pre-measurement, a classical observer can readout the qubit's state by measuring the field's state, so that the measurement accuracy depends on the overlap between the pointer states, $\mq{e^{-i\phi/2}\alpha}{ e^{i\phi/2}\alpha}= e^{-2\nAv(1-\cos{\phi})}$, where $\nAv=\abs{\alpha}^2$. The smaller the overlap between the pointer states, the more distinguishable they will be in a classical readout and hence the more accurate the measurement of $\sigma_z$ will be. The pointer states' overlap can be made arbitrarily small by increasing the number of photons, unless $\phi=0$. Such repeatable and accurate measurements of qubits are ubiquitous in circuit QED where dispersive interactions are widely used to measure superconducting (transmon) qubits in microwave resonators~\cite{Schuster2007resolving,Devoret_review,Blais2021CQED}. We remark that in the standard situation described so far, $\comm{H_Q}{V}=0$ so that no energy exchanges between system and meter occur.

In the general case, $\comm{H_Q}{V}\neq 0$ but, since our dynamics is of the scattering type\footnote{The dynamics is of scattering type because, with extremely good precision, $V\ket{\psi(t)}\sim 0$ for $t\leq -T$ and $t\geq +T$.}, we have the following property~\cite{Book_Taylor2006Scattering}: $\comm{H_Q+H_F}{U(T,-T)}=0$. Therefore, the sum of the bare energies is conserved when considering the whole dynamics. In this case, the interaction can change the qubit's internal energy, i.e., $\Delta E_{Q}=\langle H_{Q}\rangle_{T}-\langle H_{Q}\rangle_{-T}\neq 0$ and, consequently, the field's energy has to change of an equal and opposite amount, $\Delta E_{F}=\langle H_{F}\rangle_{T}-\langle H_{F}\rangle_{-T}=-\Delta E_{Q}$. As the interaction conserves the field's number of photons, the energy change $\Delta E_{F}$ corresponds to a deformation of the scattered light spectrum: the spectrum's mean frequency shifts towards blue or red according to the sign of $\Delta E_{Q}$.

Non-commuting dispersive interactions can also be implemented in circuit QED: the meter is still the microwave field in the resonator dispersively coupled to the qubit, while the Hamiltonian $H_Q$ with $\Theta \neq 0$ can be implemented by driving the transmon qubit with a resonant classical field~\cite{Blais2004Architecture,Blais2007Processing,Gambetta2008Zeno,Boissonneault2009Dispersive,Blais2021CQED,Linpeng2023QuantumEnergetics}. The natural qubit Hamiltonian along the $z$-axis is eliminated by studying the dynamics in the rotating frame with the driving frequency. This strategy has been recently used in the experiment reported in Ref.~\cite{Linpeng2023QuantumEnergetics} to investigate energy exchanges in dispersive measurements of a qubit whose Hamiltonian is tilted by $\Theta=\pi/2$.

\subsection{Formal solution of the dynamics}

As the interaction [Eq.~\eqref{eq:InteractionHamiltonian}] conserves the photon number, it is possible to show (see Appendix~\ref{APPSec:FormalSolution}) that the formal expression of the joint qubit-field state at time $t$, in the interaction picture with respect to the qubit's bare Hamiltonian and in the frame rotating with the input field's frequency $\omega_0$, reads:
\begin{multline}
\label{eq:SolutionGeneralDynamics}
\ket{\psi(t)} = \prtqB{\beta_0 + \sum_{n=1}^\infty \frac{\beta_n}{\sqrt{n!}}\times 
\\
\times \intmp \dd{t}^n \psi_n (\vec{t}_n)\tU_n (\vec{t}_n,t)A^\dg_n (\vec{t}_n)}\ket{0_F}\ket{\psi_Q (-T)},
\end{multline}
which is the same as Eq.~\eqref{eq:InitialState} with the addition of the unitary operators\footnote{In Appendix~\ref{APPSec:FormalSolution}, the operator $\tU_n (\vec{t}_n,t)$ is given in a more general form valid for any interaction shape $f(x)$.} $\tU_n (\vec{t}_n,t)$, which are defined as follows:
\begin{equation}
\label{eq:UnitaryOperator}
\tU_n (\vec{t}_n,t) = \prod_{j=1}^n e^{\frac{i t^{\uparrow}_j}{\hbar}H_0} e^{-\Theta_H (t-t_j^\uparrow)\frac{i \phi}{2}\sigma_z} e^{-\frac{i t^{\uparrow}_j}{\hbar}H_0},
\end{equation}
where the times $t_j^\uparrow$ are an increasingly ordered permutation of the times $t_j$ ($t_1^\uparrow$ is the lowest of the $t_j$, and $t_n^\uparrow$ is the highest) and the Heaviside function $\Theta_H$ is needed so that, if $t_j > t$, the interaction does not take place. The factor $1/\sqrt{n!}$ in Eq.~\eqref{eq:SolutionGeneralDynamics} erases all cases in which the same dynamics is counted multiple times and stems directly from the indistinguishability property of bosons. 

\begin{figure}[t]
	\centering
	\includegraphics[width=0.48\textwidth]{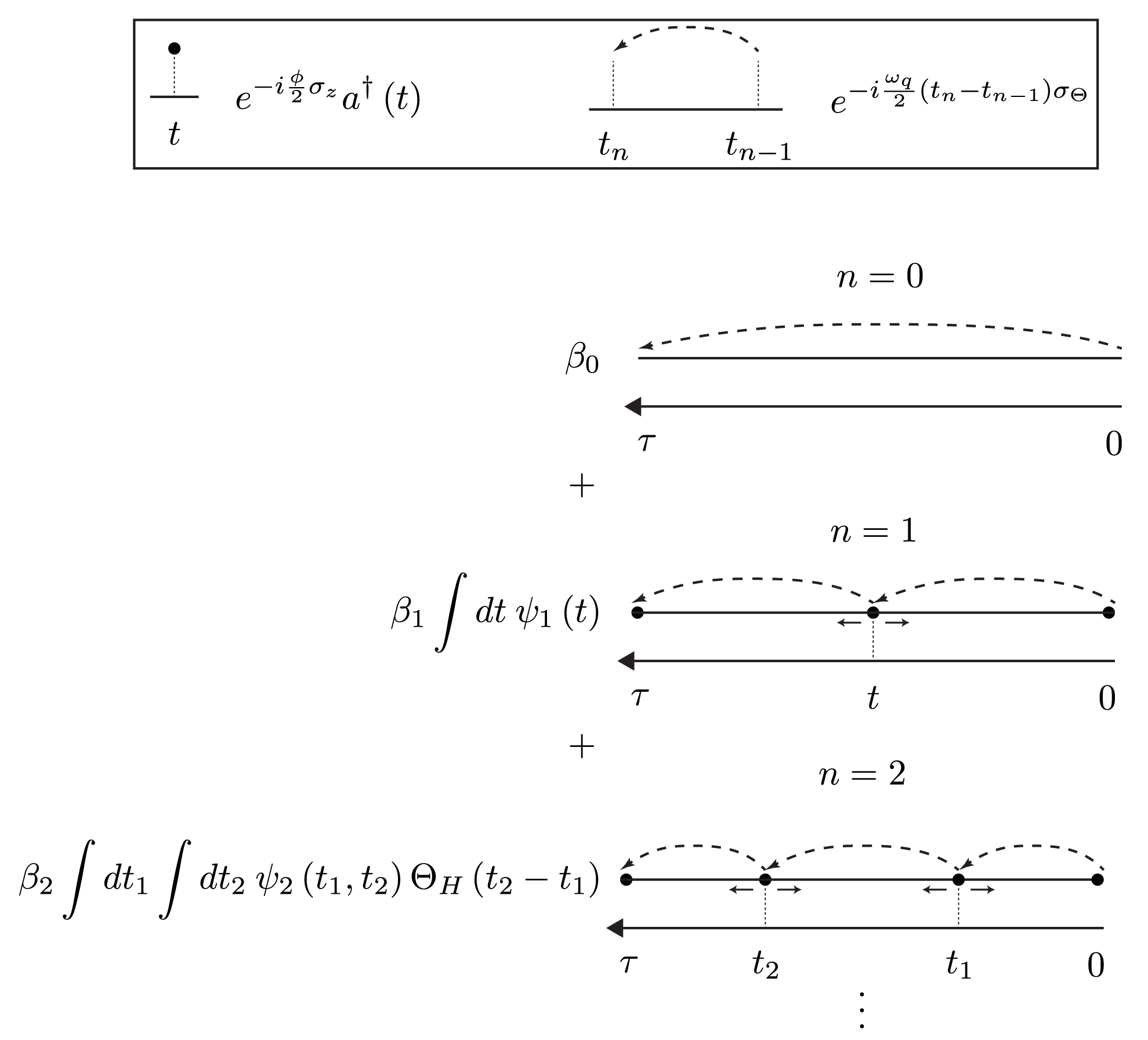}
	\caption{Pictorial representation of the solution of the joint system dynamics leading to Eq.~\eqref{eq:SolutionGeneralDynamics}. For each photon number subspace, the dynamics can be seen as a collection of sequences of free evolution and instantaneous interactions (or collisions). The amplitudes of these sequences correspond to the wavepacket wavefunction $\psi_n (\vec{t}_n)$ and they have to be summed accordingly. Notice that, in the picture, the bosonic nature of the propagating field is accounted for by the Heaviside function. This is also the reason why the term $1/\sqrt{n!}$ does not appear in the picture.}
\label{fig:SolutionCollisionScheme}
\end{figure}

The dynamics can be understood through the pictorial representation shown in Fig.~\ref{fig:SolutionCollisionScheme} whose interpretation is the following: for each photon subspace with fixed $n$, the photons collide with the qubit at times $t_1,t_2,\dots,t_n$ with $t_{j} \geq t_{j-1}$. These collisions are instantaneous, but the qubit undergoes free evolution between them, and the probability amplitudes of these processes have to be summed over all the possible $n$-tuples of times.

\section{Analysis of the field's spectral deformation \label{Sec:DynamicsSOlution}}

\subsection{Single photon pulses\label{Sec:SinglePhotonPulses}}

Here we explicitly solve the dynamics of the joint system when the initial state of the field $\ket{\psi_{F}(-T)}$ is a single-photon wavepacket: $\ket{\psi_{F}(-T)}= \intmp \dd{t} \psi_1 (t) \ad(t) \ket{0_F}$ with $\psi_{1}(t)$ being an arbitrary (normalized) temporal profile. The initial state of the joint system is defined to be $\ket{\psi (-T)}=\ket{\psi_{Q}(-T)}\ket{\psi_{F}(-T)}$, with $\ket{\psi_{Q}(-T)}= b_g \ket{g_\Theta} + b_e \ket{e_\Theta}$, where $\ket{g_\Theta}$ and $\ket{e_\Theta}$ are, respectively, the ground and excited state of $H_Q$. The explicit wavefunction (the qubit's interaction picture) arising from Eq.~\eqref{eq:SolutionGeneralDynamics} for $\beta_1=1$ is derived in Appendix~\ref{APPSec:OnePhotonSolution}. In the long time limit, when the field's scattering is complete, it reads:
\begin{multline}
\label{eq:Onep_solution}
\ket{\psi (T)} = 
\prtq{b_g I_{gg} \ket{g_\Theta} + b_e I_{ee} \ket{e_\Theta}}\ket{\psi_{F}(0)} + \\ +
b_g I_{ge}\ket{e_\Theta}\ket{\psi_{F,ge}} + b_e I_{eg}\ket{g_\Theta}\ket{\psi_{F,eg}},
\end{multline}
where $I_{\epsilon \varepsilon} \equiv \mel{\epsilon_\Theta}{e^{- i (\phi/2)\sigma_z}}{\varepsilon_\Theta}$ with $\epsilon,\varepsilon \in \lbrace e,g\rbrace$. The explicit values are
\begin{equations}
\label{eq:OnePhotonAmplitudes}
I_{gg} &= I_{ee}^* = \cos(\phi/2)+ i \cos(\Theta)\sin(\phi/2),
\\
I_{ge} &= I_{eg} = i \sin(\Theta)\sin(\phi/2).
\end{equations}
Finally, the field's distorted states are given by
\begin{equations}
\label{eq:OnePhotonFieldsTimeDomain}
\ket{\psi_{F,eg}} &= \intmp \dd{t} \psi_1 (t) e^{-i \omega_q t} \ad(t) \ket{0_F},
\\
\ket{\psi_{F,ge}} &= \intmp \dd{t} \psi_1 (t) e^{+i \omega_q t} \ad(t) \ket{0_F}.
\end{equations}

By re-writing these states in the frequency domain we can see that the phases correspond to a shift of the light's spectrum:
\begin{equations}
\label{eq:OnePhotonFieldsFrequencyDomain}
\ket{\psi_{F,eg}} &= \intmp \dd{\omega} \tpsi_1 (\omega-\omega_q) \ad(\omega) \ket{0_F},
\\
\ket{\psi_{F,ge}} &= \intmp \dd{\omega} \tpsi_1 (\omega+\omega_q) \ad(\omega) \ket{0_F},
\end{equations}
where we used the equality $\ad(t) = (1/\sqrt{2 \pi}) \intmp \dd{\omega} e^{+i \omega t}\ad (\omega)$ [cf. Eq.~\eqref{eq:BosonicOperatorsInverseTransformation}] and defined $\tpsi_1 (\omega) \equiv (1/\sqrt{2\pi}) \intmp \dd{t} \psi_1 (t) e^{+i \omega t}$, as the input field's wavefunction in the frequency domain.

Eq.~\eqref{eq:Onep_solution} shows that, as expected, the scattering with a single-photon pulse can change the qubit's energy. Since the interaction [Eq.~\eqref{eq:InteractionHamiltonian}] is such that $V\ket{\psi(\pm T)}=0$, the qubit's energy change must be compensated by the field's energy change. Equations~\eqref{eq:OnePhotonFieldsTimeDomain} and~\eqref{eq:OnePhotonFieldsFrequencyDomain} reveal the microscopic mechanism underlying this energy balance: the field's wavefunctions associated with an increase, resp. decrease, of the qubit's energy acquire a time dependent phase corresponding to a frequency red-shift, resp. blue-shift. Hence, Eq.~\eqref{eq:Onep_solution} features the superposition of the three possible system's trajectories: 
\begin{enumerate}
    \item The qubit's state does not change and the field's state remains $\ket{\psi_F (0)}$.
    \item The qubit's state changes increasing its energy and the field ends in the red-shifted state $\ket{\psi_{F,ge}}$.
    \item The qubit's state changes losing energy and the field ends in the blue-shifted state $\ket{\psi_{F,eg}}$. 
\end{enumerate}
Let us notice that when $\{\Theta,\phi\}=\{\pi/2,\pi\}$ we have that $I_{gg} = I_{ee} = 0$ and $I_{ge} = I_{eg} = i$; only the trajectories featuring the qubit's excitation or de-excitation have non-zero amplitude. Indeed, in the regime of $\pi$-per-photon interaction ($\phi=\pi$), any qubit-photon interaction leads to a $\pi$-rotation around the $z$-axis of the qubit's Bloch sphere; when the qubit tilt is maximal $(\Theta=\pi/2)$, this $\pi$-rotation swaps ground and excited states.

\begin{figure}
\centering
\includegraphics[width=0.48\textwidth]{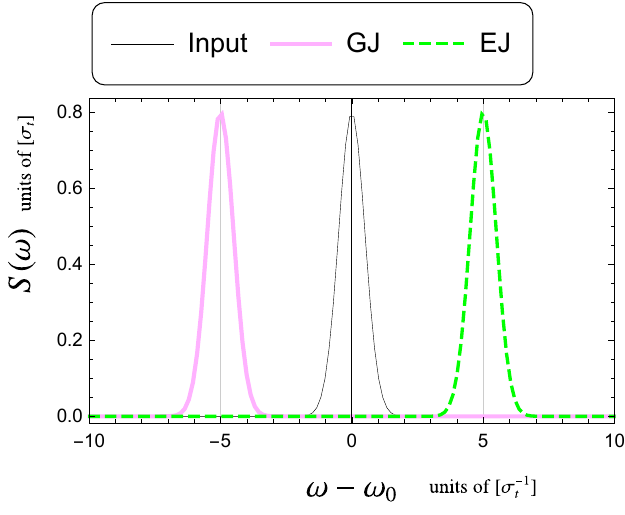}
\caption{Spectral components of the possible field's final states for a single-photon pulse. The field's initial state has central frequency $\omega_0$ and a Gaussian profile of time-dispersion $\sigma_t$ ( see Eq.~\eqref{eq:GaussianEnvelope}). In this case, $\sigma_t \omega_q = 5$.
For better comparison with the plots of Sec.~\ref{Sec:MultiPhotonPulses}, the plot already uses the nomenclature introduced later. For reference with quantities of this section: $\rm{Input} \leftrightarrow \abs{\tpsi_1 (\omega)}^2$, $\rm{GJ} \leftrightarrow \abs{\tpsi_1 (\omega+\omega_q)}^2$, and $\rm{EJ} \leftrightarrow \abs{\tpsi_1 (\omega-\omega_q)}^2$.}
\label{fig:SpectrumShiftOnePhoton}
\end{figure}

Figure~\ref{fig:SpectrumShiftOnePhoton} shows the spectra related to the three states appearing in Eq.~\eqref{eq:Onep_solution}. The plot is realized by considering an input pulse with the following Gaussian profile:
\begin{equation}
\label{eq:GaussianEnvelope}
\psi_{\rm Env} (t) = \frac{1}{(2\pi)^{1/4}\sqrt{\sigma_t}}\exp[-\prt{\frac{t}{2 \sigma_t}}^2].
\end{equation}

The spectrum $S_{\rm out}(\omega)$ of the scattered field can be calculated easily by computing $S_{\rm out}(\omega) = \ev{\ad (\omega) a(\omega)}{\Psi_{f}}$. The result is
\begin{multline}
\label{eq:OnePhotonOutputSpectrum}
S_{\rm out} (\omega) = \abs{I_{gg}}^2 S_{\rm in} (\omega) 
+\\
\abs{I_{ge}}^2 \prtq{\abs{b_g}^2 S_{\rm in}(\omega + \omega_q) + \abs{b_e}^2 S_{\rm in}(\omega - \omega_q)} 
+ 2\Re \prtgB{ \\
b_g^* b_e I_{eg} I_{gg}^* \prtq{\tpsi_1^* (\omega) \tpsi_1 (\omega - \omega_q)-\tpsi_1^* (\omega + \omega_q) \tpsi_1 (\omega)}},
\end{multline}
where we exploited the fact that $I_{gg}=I_{ee}^*$, $I_{ge}=I_{eg}$, $I_{ge}^* = - I_{ge}$, and $S_{\rm in} (\omega) = \abs{\tpsi_1 (\omega)}^2$. 
Let us denote by $\sigma_t$ the temporal dispersion of the pulse (i.e., effectively, its temporal length) and by $\sigma_\omega$ its frequency dispersion, recalling that they are linked by the Heisenberg uncertainty principle $\sigma_t \sigma_\omega \geq 1/2$. When $\omega_q \gtrsim \sigma_\omega$, the spectrum presents a three-peak structure. 
Let us notice that, for Gaussian pulses, for which $\sigma_t \sigma_\omega \sim 1$, the condition for the three-peak structure becomes $\sigma_t \omega_q \gtrsim 1$ . 

The first three terms of the spectrum in Eq.~\eqref{eq:OnePhotonOutputSpectrum} come from the three possible trajectories (1-3), while the last term features an interference between shifted and non-shifted wavefunctions. When $\sigma_\omega \ll \omega_q$, the interference term vanishes and the three-peak structure becomes clearly visible. In other words, the three-peak structure becomes clearly visible when it is possible, by measuring the photon frequency in the output, to infer with great accuracy which path the qubit followed (jump $\ket{g_\Theta} \rightarrow \ket{e_\Theta}$, jump $\ket{e_\Theta} \rightarrow \ket{g_\Theta}$, or no jump), i.e., the spectrum presents a clear three-peak structure when it carries which-path information.

Importantly, the three peaks appearing in the output spectrum are not due to processes akin to those generating the Mollow triplet~\cite{Book_Breuer2002,Book_Walls2007}. In that case, the spectral deformation arise from the exchange of excitations between the qubit and the resonant field driving it. As a consequence, the position of the Mollow triplet peaks depends on the intensity of the driving. In our model, on the contrary, the qubit-field interaction is dispersive so that no excitations are exchanged between field and qubit, and the side peaks position depends on the qubit's bare frequency and not on any of the pulse's properties. We will see that this property is maintained, under certain conditions, even when the probe pulse contains more than one photon.

The case in which $\{\Theta,\phi\}=\{\pi/2,\pi\}$ deserves a special mention as in this case $\abs{I_{gg}}=0$ and $\abs{I_{ge}} = 1$ so that the output spectrum is
\begin{equation}
\label{eq:OnePhotonOutputSpectrumMagicValues}
S_{\rm out} (\omega)\vert_{\{\Theta,\phi\}=\{\frac{\pi}{2},\pi\}} = \abs{b_g}^2 S_{\rm in}(\omega + \omega_q) + \abs{b_e}^2 S_{\rm in}(\omega - \omega_q),
\end{equation}
which presents a two peak structure regardless of the length of the pulse. Indeed, the two peaks are separated and therefore clearly identifiable as two separate peaks when $\sigma_\omega \ll \omega_q$.

\subsection{Multi-photon pulses\label{Sec:MultiPhotonPulses}}

Here we consider the case where the field is a multi-photon pulse and display the mechanism underlying the appearance, in the spectrum of the scattered field, of peaks that can be located in $\omega_0$ and/or $\omega_0 \pm \omega_q$. In this subsection, we will consider Gaussian pulses for which frequency and time dispersion are linked by $\sigma_t \sigma_\omega \sim 1$.

We start by considering pulses with exactly $n$ photons. Then, we extend our reasoning to pulses containing a superposition of states with different photon numbers, such as the coherent states considered in Sec.~\ref{sec:CoherentPulses}.
Let us start from a physical consideration following the schematics in Fig.~\ref{fig:SolutionCollisionScheme} and the analytical results obtained on the single-photon pulses. Equation~\eqref{eq:OnePhotonOutputSpectrum} shows that when a single photon interacts with the qubit its final spectrum contains in general three distinguishable peaks if $\sigma_t \omega_q \gtrsim 1$. This latter can be interpreted as a condition on the effective light-matter interaction time: the photon should interact with the qubit for a time $\sigma_t$ long enough to \enquote{measure} the qubit's bare dynamics whose timescale is $1/\omega_q$. For a pulse of $n$ photons, we can estimate an "average time-dispersion per photon" as $\sigma_t/n$, where $\sigma_t$ is the time-dispersion of the pulse. Fig.~\ref{fig:SolutionCollisionScheme} shows that for a generic initial field's state, the system's dynamics can be interpreted as a superposition of all the possible processes where each photon of the field collides with the qubit at a different time. Then we expect the side peaks in $\omega_0 \pm \omega_q$ to appear when the average interaction time per photon is long enough, i.e., $\sigma_t/n \gtrsim 1/\omega_q$ or, equivalently, $\sigma_t \omega_q \gtrsim n$. The results presented below prove that this intuition is correct.
 
For pulses containing an arbitrarily large number of photons, finding the explicit analytical expression of Eq.~\eqref{eq:SolutionGeneralDynamics} is quite complex. Here, we analyze the  final spectrum for fields prepared in Fock states with $n\leq 3$ by computing numerically the spectrum of the final field's wavefunctions associated to the different possible trajectories of the qubit.  
To denote these trajectories and the corresponding field's states, we use a sequence of $n+1$ letters where $n$ is the number of photons in the field. The first letter is either \enquote{G} or \enquote{E} and denotes whether the qubit's initial state is the ground or the excited state of $H_Q$. The other letters denote if the qubit's state flips (i.e., $\ket{e_{\Theta}/g_{\Theta}}\rightarrow \ket{g_{\Theta}/e_{\Theta}}$) or not in the interaction with the $i$-th photon of the pulse. If the qubit flips we write \enquote{J} standing for \enquote{jump}, otherwise we write \enquote{N} standing for \enquote{no-jump}. For example, for a three-photon pulse the state GJNN denotes the trajectory where the qubit started in the ground state of $H_Q$, got flipped by the interaction with the first photon, and remained in the excited state after the interaction with the other two photons of the pulse. Once the field's states are computed in the frequency domain, they can be used for the calculation of the corresponding spectral component (more details are given in Appendix~\ref{APPSec:SpectrumMultiPhoton}). Importantly, let us notice that the wavefunctions (and thus their spectrum) corresponding to the possible system's trajectories do not depend on the values of $\Theta$ and $\phi$, as the latter parameters only control the weight that every wavefunction has in the overall output state of the field. This is already visible in the single-photon wavefunctions of  Eq.~\eqref{eq:Onep_solution}, where $\Theta$ and $\phi$ only appear in the coefficients $I_{ge}$ and $I_{gg}$. Finally, it is worth noting that when $\prtg{\Theta,\phi}=\prtg{\pi/2,\pi}$ each photon interaction leads to a jump of the qubit's state so that only the states with all jumps\footnote{For example, with three photons, only the states GJJJ and EJJJ will appear in the final state of the field-qubit system.} have non-zero amplitudes in front of them at $t=T$. We will see that this, as expected, leads to spectra with only the two peaks instead of three.

\begin{figure}
	\centering
	\includegraphics[width=0.48\textwidth]{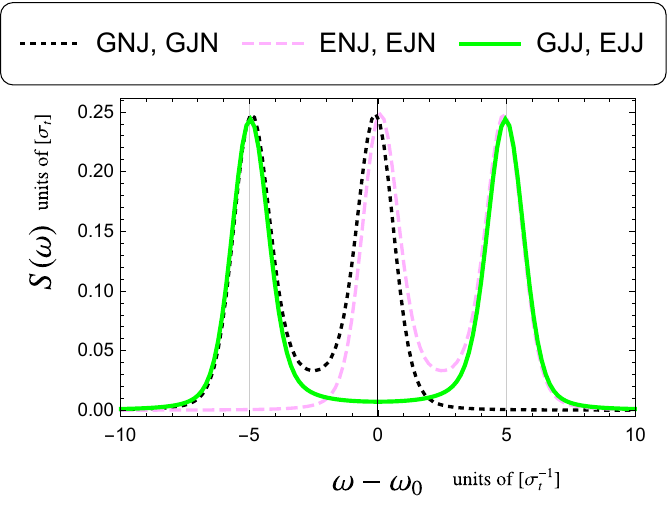}
	\caption{Spectral components of the possible field's final states for a two-photon pulse. The field's initial state has central frequency $\omega_0$ and a Gaussian profile of time-dispersion $\sigma_t$ ( see Eq.~\eqref{eq:GaussianEnvelope}). Every spectral component is normalized to one and $\sigma_t \omega_q = 5$. We recall that the above plots do not depend on either $\Theta$ or $\phi$.}
	\label{fig:SpectrumShiftTwoPhotons}
\end{figure}

\begin{figure}
	\centering
	\includegraphics[width=0.48\textwidth]{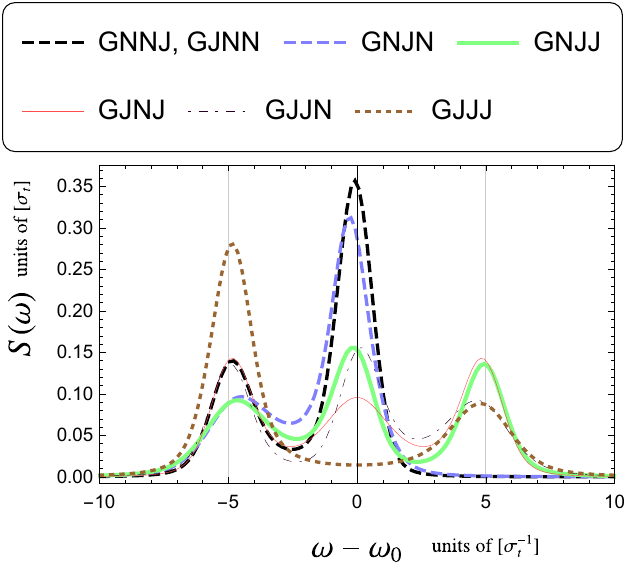}
	\caption{Spectral components of the possible field's final states for a three-photon pulse and the qubit starting in the ground state $\ket{g_{\Theta}}$.  The field's initial state has central frequency $\omega_0$ and a Gaussian profile of time-dispersion $\sigma_t$ ( see Eq.~\eqref{eq:GaussianEnvelope}). Every spectral component is normalized to one and $\sigma_t \omega_q = 5$. We recall that the above plots do not depend on either $\Theta$ or $\phi$.}
	\label{fig:SpectrumShiftThreePhotonsFromGround}
\end{figure}

\begin{figure}
	\centering
	\includegraphics[width=0.48\textwidth]{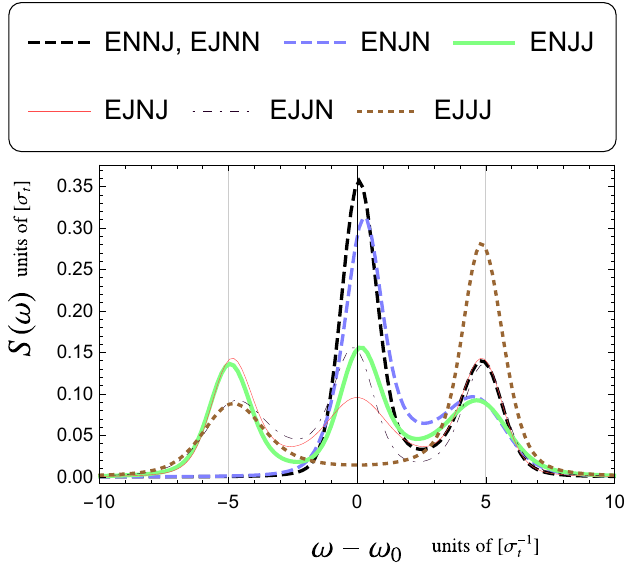}
	\caption{Spectral components of the possible field's final states for a three-photon pulse and the qubit starts in the excited state $\ket{e_{\Theta}}$ The temporal shape of the input pulse is given in Eq.~\eqref{eq:GaussianEnvelope}. Every spectral component is normalized to one and $\sigma_t \omega_q = 5$. We recall that the above plots do not depend on either $\Theta$ or $\phi$.}
	\label{fig:SpectrumShiftThreePhotonsFromExcited}
\end{figure}

\begin{figure}
\centering
\includegraphics[width=0.48\textwidth]{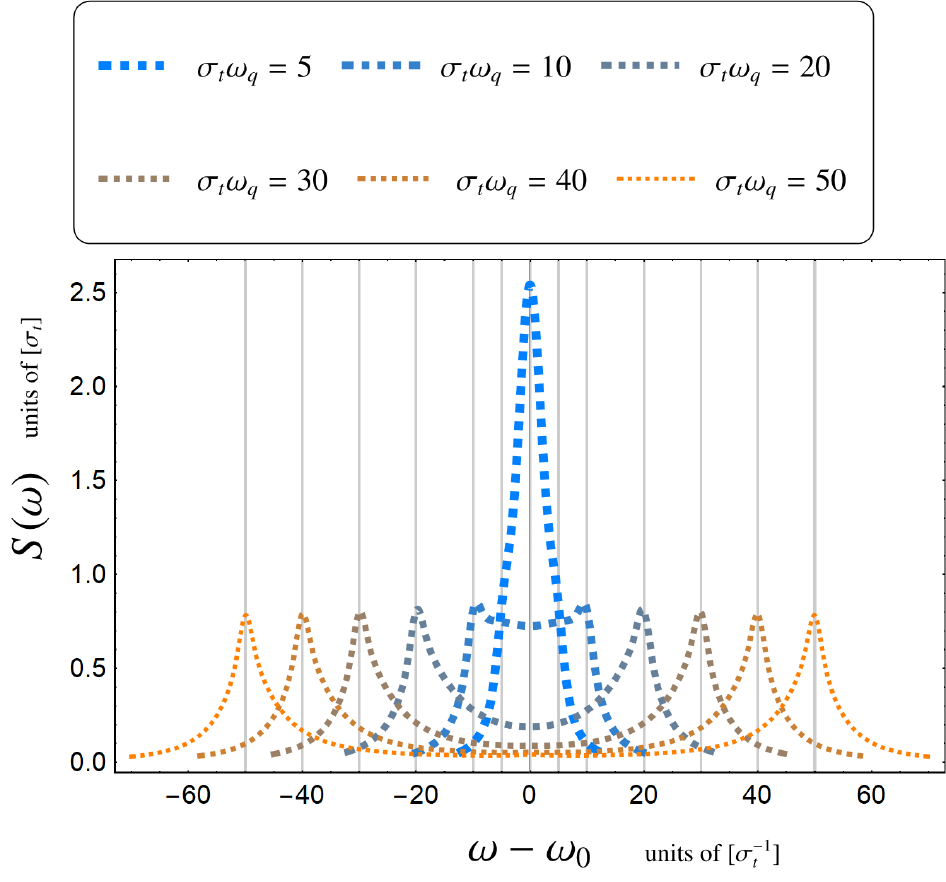}
\caption{Spectra of the scattered field when the field is prepared in a coherent pulse containing 20 photons on average varying the pulse's duration. The plots are obtained for $\Theta = \pi/2$ and $\phi=\pi$. The two-peak structure is clearly visible when $\sigma_t \omega_q \gtrsim 20$.}
\label{fig:Spectra20Photons}
\end{figure}

Fig.~\ref{fig:SpectrumShiftOnePhoton} refers to a field carrying just one photon ($n=1$). It shows the spectra of the field's wavefunctions corresponding to the trajectories EJ and GJ, respectively $\ket{\psi_{F,eg}}$ and $\ket{\psi_{F,ge}}$ of Eq.~\eqref{eq:OnePhotonFieldsFrequencyDomain}, alongside the input spectrum (GN and EN have the same spectrum of the input pulse). The plot is realized by considering an input pulse with the Gaussian profile of Eq.~\eqref{eq:GaussianEnvelope}. For input pulses carrying $n>1$ photons, we set their initial temporal profile to be $\psi_n (\vec{t}_n) = \prod_{j=1}^n \psi_{\rm Env} (t_j)$.

Fig.~\ref{fig:SpectrumShiftTwoPhotons} refers to a field carrying two photons ($n=2$) and shows the spectra of the six possible field's final states corresponding to the trajectories GNJ, GJN, ENJ, EJN, GJJ, and EJJ. 

Finally, Figs.~\ref{fig:SpectrumShiftThreePhotonsFromGround} and~\ref{fig:SpectrumShiftThreePhotonsFromExcited} refer to a field carrying three photons ($n=3$) and the qubit starting respectively in the ground and in the excited state. As it is clear by comparing the two plots, the trajectories where the qubit starts in the ground state (Fig.~\ref{fig:SpectrumShiftThreePhotonsFromGround}) give spectral contributions being the mirrorlike images of those where the qubit starts in the excited state (Fig.~\ref{fig:SpectrumShiftThreePhotonsFromExcited}). Again, as expected, all the spectral contributions have peaks around $\omega_0$ and $\omega_0 \pm \omega_q$. Interestingly, here we can see that the shapes of the spectral components do not depend uniquely on the number of qubit's jumps, but also on when they happen: the spectral component corresponding to GNNJ is different from that of GNJN; and so the spectral components coming from the trajectories GNJJ, GJNJ and GJJN are different among each other.

When the input field is a superposition of multi-photon pulses with different photon numbers $n$, the final spectrum can be computed as the weighted sum of the spectrum coming from each multi-photon pulse separately. Then, when each of these spectral components presents the peaks in the same three positions ($\omega_0$, and $\omega_0 \pm \omega_q$), the resulting spectrum will also have peaks in these positions. Following the qualitative analysis presented at the beginning of this section, we expect that the condition for having well-defined peaks in $\omega_0 \pm \omega_q$ is $\sigma_t \omega_q \gtrsim \nAv$, where $\nAv$ is the average number of photons in the pulse. Fig.~\ref{fig:Spectra20Photons} shows the output spectra in the case of a coherent input pulse with\footnote{Obtaining such spectra with the methods of Sec.~\ref{Sec:MultiPhotonPulses} would be computationally prohibitive, We explain how to obtain the output spectra with coherent input pulses in Sec.~\ref{sec:CoherentPulses}.} $\nAv=20$, $\prtg{\Theta,\phi} = \prtg{\pi/2,\pi}$, changing the pulse's duration and starting from the qubit's state $\ket{g_z}$. As $\prtg{\Theta,\phi} = \prtg{\pi/2,\pi}$, the qubit's state flips at each photon-qubit interaction. As expected, we observe a two-peak structure (no central peak) for $\sigma_t \omega_q \gtrsim 20$. 
On the other hand, for $\sigma_t \omega_q = 10$, the spectrum has a pronounced plateau between the peaks in $\omega_0 \pm \omega_q$, and for  $\sigma_t \omega_q = 5$ it exhibits a big central peak in $\omega_0$.

\section{Coherent pulses\label{sec:CoherentPulses}}

In this section, we numerically compute the spectrum of the scattered field assuming that the input field is a coherent pulse with $\nAv = 4$ photons on average. We chose this value as, for $\nAv = 4$ and $\Theta=0$, the pointer states' overlap, $\mq{e^{-i\phi/2}\alpha}{ e^{i\phi/2}\alpha}= e^{-2\nAv(1-\cos{\phi})}$, is very small for any value of $\phi \in [\pi/3,\pi]$. This means that, at $\Theta=0$, a coherent pulse with $\nAv = 4$ is a good quantum meter for $\sigma_z$, i.e., it leads to distinguishable pointer states. We stress that the formulas given in this section are valid only when the probe pulse is a coherent state. These formulas provide access to fast numerical computations for the qubit's dynamics and the output spectra.

In order to compute the spectrum of the scattered pulse, we first derive the qubit's master equation (details are given in Appendix~\ref{APPSec:CollisionModelherentState}). The qubit's master equation\footnote{Despite the appearances, the master equation is in Lindblad form because the jump operator is unitary.} reads
\begin{equation}
\label{eq:QubitMasterEquationCoherentPulse}
\dot{\rho} (t) = i \frac{\omega_q}{2}\comm{\rho(t)}{\sigt}+ \abs{\alpha(t)}^2 \prtq{e^{- i \frac{\phi}{2}\sigma_z} \rho(t) e^{+i \frac{\phi}{2}\sigma_z} - \rho(t)},
\end{equation}
where $\rho (t)$ is the qubit density matrix and $\alpha(t)$ is the wavefunction associated to the coherent pulse~\cite{Book_Loudon2000}. In our case the pulse has a Gaussian shape:
\begin{equation}
\label{eq:GaussianEnvelopeCoherent}
\alpha (t) = \frac{\sqrt{\nAv}}{(2\pi)^{1/4}\sqrt{\sigma_t}}\exp[-\prt{\frac{t}{2 \sigma_t}}^2],
\end{equation}
with $\int \dd{t} \abs{\alpha(t)}^2 = \nAv$ being the average photon number.
From the master equation we can compute the autocorrelation function of the output pulse as follows (see Appendix~\ref{APPSec:CollisionModelherentState}):
\begin{equation}
g(s,t) = \alpha^* (s) \alpha(t) \Tr \prtg{e^{+i \frac{\phi}{2}\sigma_z} \mcE_{t,s} \prtq{e^{- i \frac{\phi}{2}\sigma_z}\rho(t)}},
\end{equation}
where the super-operator $\mcE_{t,s}$ evolves its argument from $t$ to $s$ according to Eq.~\eqref{eq:QubitMasterEquationCoherentPulse}. This expression is valid for $s \geq t$, and for $s<t$ we exploit the fact that $g(s,t)^* = g(t,s)$. The spectrum of the scattered field is obtained by computing the integral:
\begin{equation}
\label{eq:SpectrumCoherentPulses}
S(\omega) = \frac{1}{2\pi} \intmp \dd{t} \dd{s} e^{-i\omega(s-t)}g(s,t).
\end{equation}

\begin{figure}[t]
\centering
\includegraphics[width=0.48\textwidth]{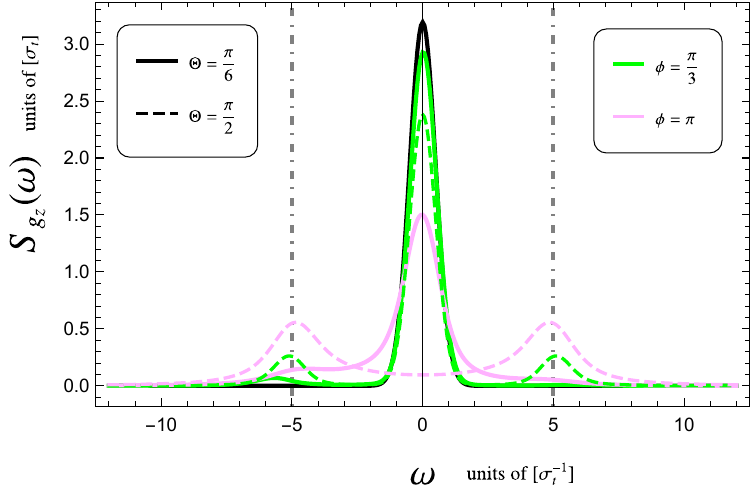}
\caption{Spectrum of the scattered pulse for $\sigma_t \omega_q = 5$. The initial state of the qubit is $\exp[i (t/\hbar) H_Q]\ket{g_z}$. The black thick line is the input spectrum, and its central frequency $\omega_0$ is set as the zero of the axis. We can observe the presence of side peaks around $\pm \sigma_t \omega_q$.}
\label{fig:specGroundZ}
\end{figure}

\begin{figure}[t]
\centering
\includegraphics[width=0.48\textwidth]{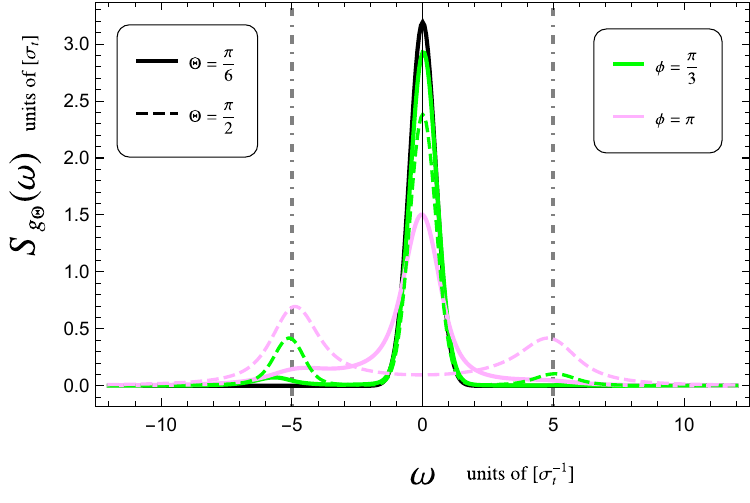}
\caption{Spectrum of the output pulse for $\sigma_t \omega_q = 5$. The initial state of the qubit is $\ket{g_\Theta}$. The black thick line is the input spectrum, and its central frequency $\omega_0$ is set as the zero of the axis. We can observe the presence of side peaks around $\pm \sigma_t \omega_q$.}
\label{fig:specGroundTheta}
\end{figure}

Figures~\ref{fig:specGroundZ} and~\ref{fig:specGroundTheta} show the spectra of the scattered fields for long pulses ($\sigma_t \omega_q = 5$) and different values of $\Theta$ and $\phi$ starting from the states\footnote{This state is chosen so that the qubit would be in state $\ket{g_z}$ at time $t=0$ if it were not for the pulse. The reason we do this is that it improves the measurement performance, as argued later in the text. See also Fig.~\ref{fig:FigWaveguide}} $\exp[i (T/\hbar) H_Q]\ket{g_z}$ and $\ket{g_{\Theta}}$, respectively. They also show, for comparison, the input spectrum as a black continuous line. We observe the occurrence of peaks around the three values $\omega_0=0$, and $\omega_0 \pm \omega_q$ as explained in the previous section. Moreover, we see that for $\{\Theta,\phi \}=\{\pi/2,\pi \}$ we do not have the central peak, as expected.
We see that, when $\Theta = \pi/2$, the spectra remain symmetric around $\omega_0=0$, while they become asymmetric when $\Theta = \pi/6$. This is explained by the fact that, at $\Theta = \pi/2$, the evolution conserves the average qubit's energy: the initial qubit's state, $\exp[i(t/\hbar)H_Q]\ket{g_z}$, and the final one, the completely mixed state\footnote{Despite the fact that we simulate a pre-measurement dynamics, the final qubit state is practically completely mixed because the pulse is long and, therefore, the measurement has bad performances.}, have the same energy. Consequently the spectrum of the scattered field remains symmetric around $\omega_0=0$ and provides the same average energy as the initial field's state. This is not the case when $\Theta=\pi/6$. In Fig.~\ref{fig:specGroundTheta}, the initial qubit state is the ground state of the Hamiltonian $H_Q$. In this case, the qubit always acquires energy from the pulse and this is why the spectra present an average red-shift in all cases.

We have seen that spectral deformations are visible only when $\sigma_\omega/\omega_q \ll 1$. According to the WAY theorem~\cite{Ozawa2002Conservation,Loveridge2011Measurement}, in this case, the qubit's measurement cannot be projective, i.e., accurate and repeatable. In Appendix~\ref{APPsec:WAYTheorem}, we give more details about the WAY theorem and how its statement applies to our system. However, a thorough quantitative benchmark of the quantum measurement in terms of accuracy, repeatability, and signal-to-noise ratio is beyond the scope of the present paper.

 \section{Conclusions\label{Sec:Conclusions}}

We investigated an autonomous pre-measurement dynamics where a qubit interacts dispersively with a light pulse propagating in a one-dimensional waveguide. We considered different statistics of the field, as well as different cases where the qubit's Hamiltonian does not commute with the measured observable ($\sigma_z$). In this scenario, the spectrum of the scattered field gets deformed by the interaction leading to a change of the field's average energy balancing the qubit's energy change while keeping the excitation number constant. We solved the qubit-field dynamics and explored the mechanism underlying the field's spectral deformation. In particular, we identified configurations of pulses' duration and photon number such that the spectrum of the scattered field has three peaks, one centered around the input field's carrier frequency and the other two shifted of $\pm \omega_q$ with respect to it, $\omega_q$ being the qubit's frequency, in agreement with the behavior observed in Ref.~\cite{Linpeng2023QuantumEnergetics}. Finally, we numerically tested this prediction by using a coherent field as quantum meter.

\section*{Acknowledgements}

This work was supported by the John Templeton Foundation (Grant No. 61835).
M. M. acknowledges the support by PNRR MUR project PE00000023-NQSTI.
N. P. acknowledges the support by PNRR NQSTI Spoke 1 “Foundations and architectures for quantum information processing and communication” CUP: J93C22001510006.
A.A. acknowledges the National Research Foundation, Singapore and A*STAR under its CQT Bridging Grant, the Plan France 2030 through the project NISQ2LSQ ANR-22-PETQ-0006 and through the project OQuLus ANR-23-PETQ-0013, the ANR Research Collaborative Project ``QuRes" (Grant ANR-PRC-CES47-0019) and “Qu-DICE” (Grant ANR-PRC-CES47), the Foundational Questions Institute Fund (Grants No. FQXi-IAF19-01 and FQXi-IAF19-05). The authors are very grateful to the whole quantum energy team for inspiring discussions, especially Patrice Camati, Robert Whitney, Samyak Prasad and Lea Bresque.

\clearpage
\onecolumngrid

\appendix

\section{Free Hamiltonian of the waveguide field\label{APPSec:FreeHamiltonian}}

In this appendix, for convenience of the reader and following Ref.~\cite{Shen2009Theory}, we report the calculation needed to write the Hamiltonian of Eq.~\eqref{eq:WaveguideHamiltonian} in the form given in Eq.~\eqref{eq:WaveguideHamiltonianSpaceRepresentation}. Then, we introduce the general form for the field wavefunction needed to find the general solution of the dynamics in appendix~\ref{APPSec:FormalSolution}.

We are interested in pulses which are narrow in frequency and centered around $\omega_0$, i.e., $\Delta \omega/\omega_0 \ll 1$, where $\Delta \omega$ is the frequency dispersion of the pulse. We then write $\omega_k \simeq \omega_0 + v_0 k$, where we write just $k$ and not $k_R$ as in Ref.~\cite{Shen2009Theory} because the waveguide we consider is chiral and we consider only right-moving pulses. Finally, we define the creation and annihilation operators at position $x$ as follows:
\begin{equation}
a_k \equiv \frac{1}{\sqrt{2\pi}}\intmp \dd{x} a (x) e^{-ikx},
\quad
a^\dg_k \equiv \frac{1}{\sqrt{2\pi}}\intmp \dd{x} a^\dg (x) e^{+ikx},
\end{equation}
which in turn imply that
\begin{equation}
\label{APPeq:BosonicOperatorsInverseTransformation}
a (x) \equiv \frac{1}{\sqrt{2\pi}}\intmp \dd{k} a_k e^{+ikx},
\quad
a^\dg (x) \equiv \frac{1}{\sqrt{2\pi}}\intmp \dd{k} \ad_k e^{-ikx},
\quad
\comm{a(x)}{\ad (y)} = \delta (x-y).
\end{equation}
We can now write:
\begin{multline}
H_{F}/\hbar 
= \intmp \dd{x} \dd{y} a^\dg (x) a (y) \intmp \frac{\dd{k}}{2 \pi} \prt{\omega_0 + v_0 k} e^{+ik(x-y)}
\\
= \intmp \dd{x} \dd{y} a^\dg (x) a (y) \prt{\omega_0 - i v_0 \partial_x}\intmp \frac{\dd{k}}{2 \pi}  e^{+ik(x-y)}
= \intmp \dd{x} \dd{y} a^\dg (x) a (y) \prt{\omega_0 - i v_0 \partial_x}\delta(x-y)
\\
= \omega_0 \intmp \dd{x} a^\dg (x) a (x)
- i \hbar v_0 \intmp \dd{x} a^\dg (x) \partial_x a (x),
\end{multline}
which corresponds to Eq.~\eqref{eq:WaveguideHamiltonianSpaceRepresentation}.

An arbitrary state of the electromagnetic field in the waveguide can be written as follows
\begin{equation}
\label{APPeq:FreeFieldNumberDecomposition}
\ket{\psi_F (t)} = 
\sum_n \beta_n \ket{n_F (t)}
= \prtq{\beta_0 + \sum_{n=1}^{\infty} \frac{\beta_n}{\sqrt{n!}} \intmp \dd{x^n} \phi_n (\vec{x}_n,t) A^\dg_n (\vec{x}_n)}\ket{0_F},
\end{equation}
where $\dd{x^n} \equiv \dd{x_1} \dd{x_2} \dots \dd{x_n}$, $\vec{x}_n \equiv x_1,x_2,\dots,x_n$, and $A^\dg_n (\vec{x}_n) \equiv \ad(x_1) \ad(x_2) \dots \ad(x_n)$.
Eq.~\eqref{APPeq:FreeFieldNumberDecomposition} is a decomposition of the field in subspaces with fixed numbers of photons, i.e., introducing the number operator $\hn \equiv \intmp \dd{k} \ad_k a_k$, we have that $\hn \ket{n_F (t)} = n \ket{n_F (t)}$ so that $\ket{n_F (t)}$ is the projection of $\ket{\psi_F (t)}$ on a specific photon-number subspace. $\ket{0_F}$ is the vacuum state and $\sum_n \abs{\beta_n}^2 = 1$ so that for any photon-number we have written a proper normalized state because we also assume that $\intmp \dd{x^n} \abs{\phi_n (\vec{x}_n,t)}^2 = 1$. Moreover, the $\phi_n (\vec{x}_n,t)$ are invariant to permutations of the $x_i$ because we are describing a bosonic field. Finally, the factor $1/\sqrt{n!}$ is needed to normalize the state: for any number of photons, we do not want to count multiple times the same state due to the indistinguishability of bosons. Formally, according to Wick's theorem, we have that
\begin{equation}
\label{APPeq:VacuumAverageAnnihilationCreationSpace}
\ev{A_n (\vec{y}_n) A^\dg_n (\vec{x}_n)}{0} = \sum_{\prtg{P(i)}} \prod_{i} \delta (y_i - x_{P(i)}),
\implies 
\beta_n \phi_n (\vec{x}_n,t) = \frac{1}{\sqrt{n!}}\mel{0}{A^\dg_n (\vec{x}_n)}{\psi_F (t)},
\end{equation}
where $\prtg{P (i)}$ are the permutations of the index $i$. The number of permutations is equal to $n!$, thus proving that $\sqrt{n!}$ is the correct normalization factor.

\clearpage

\section{Formal solution of the dynamics\label{APPSec:FormalSolution}}

In this section, we present the formal solution to the dynamics determined by the Hamiltonian $H = H_F + H_0 + V$, where $H_F$ is given in Eq.~\eqref{eq:WaveguideHamiltonianSpaceRepresentation}, $V$ is given in Eq.~\eqref{eq:InteractionHamiltonian} and $H_0$ is a generic Hamiltonian for the fixed scatterer (system $S$). Notice that, in the treatment presented here, also the Hamiltonian $H_1$ appearing in Eq.~\eqref{eq:InteractionHamiltonian} can be completely arbitrary.
Notice that, in this Appendix, the initial time is taken to be $t=0$ instead of $t=-T$ in order to lighten the notation. The formulas reported in the main text are written considering the initial time to be $t=-T$.

Since $V$ does not change the number of photons, if the initial state of the whole system is a product state $\ket{\psi_F (0)}\otimes \ket{\psi_S (0)}$, we can study the dynamics as divided in each photon number subspace. Therefore, we write [see Eq.~\eqref{APPeq:FreeFieldNumberDecomposition}]
\begin{equation}
\label{APPeq:PhotonNumberDecomposition}
\ket{\psi (t)} =  \prtq{\beta_0 e^{-\frac{i t}{\hbar}H_0} + \sum_{n=1}^\infty \frac{\beta_n}{\sqrt{n!}} \intmp \dd{x^n} \phi_n (\vec{x}_n,t) \tV_n(\vec{x}_n,t) A^\dg_n (\vec{x}_n)}\ket{0}\ket{\psi_S (0)},
\end{equation}
where $\tV_n (\vec{x}_n,t)$ is a unitary operator on the Hilbert space of system $S$ with $\tV_n (\vec{x}_n,0) = \Id$.
Deriving the state with respect to time in a given $n$-photon subspace as done in the \Schr equation leads to
\begin{equation}
i \hbar \partial_t \ket{\psi (t)} =  \frac{\beta_n}{\sqrt{n!}} \intmp \dd{x^n} \prtq{ i \hbar \tV_n(\vec{x}_n,t) \partial_t \phi_n (\vec{x}_n,t)  + i\hbar \phi_n (\vec{x}_n,t) \partial_t \tV_n(\vec{x}_n,t)} A^\dg_n (\vec{x}_n)\ket{0}\ket{\psi_S (0)}.
\end{equation}
In the same way, the application of $H_{F}$ to the state $\ket{\psi (t)}$, now reads
\begin{equation}
H_{F} \ket{\psi (t)} = \frac{\beta_n}{\sqrt{n!}} \intmp \dd{x^n} \prtq{ \prtq{\hbar \omega_0 n - i \hbar v_0 \prt{\sum_{j=1}^n \partial_{x_j}} } \phi_n (\vec{x}_n,t) \tV_n(\vec{x}_n,t)} A^\dg_n (\vec{x}_n)\ket{0}\ket{\psi_S (0)}
\end{equation}

By making again use of the commutation relation $\comm{c (x)}{c^\dg (y)}= \delta(x-y)$ and Wick's theorem, we now compute the effect of $V$ on $\ket{\psi (t)}$, obtaining
\begin{equations}
V \ket{\psi (t)} 
&= \sum_n \frac{\beta_n}{\sqrt{n!}} \intmp \dd{y} \dd{x^n} f(y) \phi_n (\vec{x}_n,t) \prtq{H_1\tV_n (\vec{x}_n,t)} \cd (y) c (y) A^\dg_n (\vec{x}_n)\ket{0} \ket{\psi_S (0)},
\\
&= \sum_n \frac{\beta_n}{\sqrt{n!}} \intmp \dd{y} \dd{x^n} f(y) \phi_n (\vec{x}_n,t) \prtq{H_1\tV_n (\vec{x}_n,t)} \prt{\sum_{j=1}^n \delta (x_j - y)} A^\dg_n (\vec{x}_n)\ket{0}\ket{\psi_S (0)},
\\
&= \sum_n \frac{\beta_n}{\sqrt{n!}} \intmp \dd{x^n} \prt{\sum_{j=1}^n f(x_j) \phi_n (\vec{x}_n,t) \prtq{H_1\tV_n (\vec{x}_n,t)}} A^\dg_n (\vec{x}_n)\ket{0} \ket{\psi_S (0)},
\\
\end{equations}
Finally, from the \Schr equation we get, for a specific $n$-photon subspace:
\begin{equation}
\label{APPeq:PhotonNumberInteractionMotionEquation} 
i \hbar \partial_t \prtq{\tV_n(\vec{x}_n,t) \phi_n (\vec{x}_n,t)}
=
\prtq{\hbar \omega_0 n - i \hbar v_0 \prt{\sum_{j=1}^n \partial_{x_j}} + H_0 + \prt{\sum_{j=1}^n f(x_j)H_1}} \phi_n (\vec{x}_n,t) \tV_n(\vec{x}_n,t).
\end{equation}
The above equation can be simplified by going in the waveguide field rotating frame at frequency $\omega_0$. We do this as follows:
\begin{equation}
\phi_n (\vec{x}_n,t) = \psi_n (\vec{x}_n - v_0 t)e^{-i \omega_0 n t},
\qq{where}
\vec{x}_n - v_0 t \equiv x_1 -v_0 t, \dots, x_n - v_0 t.
\end{equation}
By inserting the above ansatz into Eq.~\eqref{APPeq:PhotonNumberInteractionMotionEquation} one gets
\begin{equation}
i \hbar \partial_t \tV_n(\vec{x}_n,t) = - i \hbar v_0 \prt{\sum_{j=1}^n \partial_{x_j}} \tV_n(\vec{x}_n,t) + \tH (\vec{x}_n)  \tV_n(\vec{x}_n,t),
\end{equation}
where, $\tH (\vec{x}_n) \equiv H_0 + \sum_{j=1}^n f(x_j) H_1$.
The above equation resembles one that could be derived by the so-called clock Hamiltonian~\cite{Aharonov1984Quantum,Aharonov1998Measurement,Gisin2018Quantum,Soltan2021Conservation} for many moving clocks. By writing $\tV_n(\vec{x}_n,t) = \tU_n(\vec{x}_n ; \vec{x}_n - v_0 t)$ and making the substitution $x_i \rightarrow x_i + v_0 t$ we arrive at
\begin{equation}
\label{APPeq:NonAutonomousDynamics}
i \hbar \partial_t \tU_n(\vec{x}_n + v_0 t; \vec{x}_n) =  \tH (\vec{x}_n + v_0 t)  \tU_n(\vec{x}_n + v_0 t ; \vec{x}_n),
\implies
\tU_n(\vec{x}_n + v_0 t ; \vec{x}_n) = \mcT \exp[-\frac{i}{\hbar} \int_0^t \dd{s} \tH (\vec{x}_n + v_0 s)],
\end{equation}
where $\mcT$ denotes the time-ordering operator. The above solution is, in fact, a generalization of the Clock Hamiltonian solution in the $n$-particle case. Notice that $\tH (\vec{x}_n)$ is invariant to permutations of the $x_i$ and the same follows for $\tU_n(\vec{x}_n + v_0 t ; \vec{x}_n)$.
The dynamics of the full system is therefore given by
\begin{equation}
\label{APPeq:FullDynamicsSolution}
\ket{\psi (t)} 
=
\prtq{\beta_0 e^{-(i t/\hbar)H_0} + \sum_{n=1}^{\infty} \frac{\beta_n}{\sqrt{n!}} e^{-i \omega_0 n t} \intmp \dd{x^n} \psi_n (\vec{x}_n) \tU_n(\vec{x}_n + v_0 t ; \vec{x}_n) A^\dg_n (\vec{x}_n + v_0 t)}\ket{0_F}\ket{\psi_S (0)},
\end{equation}
Notice that the above equation implies that the wavepacket is not deformed during the dynamics. This can be easily seen by the fact that functions $\abs{\psi_n (\vec{x}_n)}^2$ do not depend on time.

When $f(x) = L \delta(x)$, the unitary operator $\tU_n(\vec{x}_n + v_0 t ; \vec{x}_n)$ can be cast into the form given in Eq.~\eqref{eq:UnitaryOperator} by making the substitution $t_j = - x_j/v_0$. The time-ordering appears in Eq.~\eqref{eq:UnitaryOperator} under the form of the impact times ordering, i.e., by the ordering imposed by writing $t_j^\uparrow$. For later uses and ease of lecture, we rewrite Eq.~\eqref{APPeq:VacuumAverageAnnihilationCreationSpace} in the time-domain:
\begin{equation}
\label{APPeq:VacuumAverageAnnihilationCreationTime}
\ev{A_n (\vec{s}_n) A^\dg_n (\vec{t}_n)}{0} = \sum_{\prtg{P(i)}} \prod_{i} \delta (s_i - t_{P(i)})
\end{equation}
where $\prtg{P (i)}$ are the permutations of the index $i$ and we recall that the number of permutations is equal to $n!$.

We can easily obtain the reduced state of system $S$ at time $t$ by considering that
\begin{equation}
    \Tr_F \prtq{A^\dg_n \prt{\vec{t}_n}\dyad{0_F}A_n \prt{\vec{s}_n}} = \ev{A_n \prt{\vec{s}_n} A^\dg_n \prt{\vec{t}_n}}{0_F}.
\end{equation}
Exploiting the permutation invariance of $\psi_n (\vec{t}_n)$ and $\tU_n \prt{\vec{t}_n,t}$ we get that
\begin{equation}
\label{APPeq:ReducedStateSystemS}
\rho_S (t) = \sum_n \abs{\beta_n}^2 \intmp \dd{t}^n \abs{\psi_n (\vec{t}_n)}^2 \tU_n \prt{\vec{t}_n,t} \dyad{\psi_S (0)} \tU_n^\dg \prt{\vec{t}_n,t}.
\end{equation}

\clearpage

\section{Solution of the dynamics for one-photon pulses\label{APPSec:OnePhotonSolution}}

Here we report the full solution of the system dynamics in the case when the pulse contains exactly one photon. Notice that, in this Appendix, the initial time is taken to be $t=0$ instead of $t=-T$ in order to lighten the notation. The formulas reported in the main text are written considering the initial time to be $t=-T$.

In the interaction picture, we write the state at time $t$ as follows [see Eqs.~\eqref{eq:SolutionGeneralDynamics} and~\eqref{eq:UnitaryOperator}]
\begin{equation}
\label{APPeq:SolutionGeneralZeroPlusOne}
\ket{\psi(t)} = \prtqB{\int_{-\infty}^{t} \dd{t_1} \psi_1 (t_1)\tU_1 (t_1) \ad (t_1) + \int_{t}^{+\infty} \dd{t_1} \psi_1 (t_1) \ad (t_1)}\ket{0_F}\ket{\psi_Q (0)},
\end{equation}
where we wrote $\tU_1 (t_1)$ instead of $\tU_1 (t_1,t)$ because we dealt with the Heaviside function by separating the integral in two parts. By substituting in the above expression the actual Hamiltonians involved in the dynamics, we get that
\begin{equation}
\ket{\psi_I (t)} = 
\ket{\psi_{F}'' (t)} \ket{\psi_Q (0)} 
+
\prtgB{\prtqB{b_g I_{gg} \ket{g_\Theta} + b_e I_{ee} \ket{e_\Theta}}\ket{\psi_{F}' (t)} +
b_g I_{ge}\ket{e_\Theta}\ket{\psi_{F,ge} (t)} + b_e I_{eg}\ket{g_\Theta}\ket{\psi_{F,eg} (t)}},
\end{equation}
where $I_{\epsilon \varepsilon} \equiv \mel{\epsilon_\Theta}{e^{- i (\phi/2)\sigma_z}}{\varepsilon_\Theta}$ with $\epsilon,\varepsilon \in \lbrace e,g\rbrace$, their explicit values being
\begin{equation}
\label{APPeq:OnePhotonAmplitudes}
I_{gg} = I_{ee}^* = \cos(\phi/2)+ i \cos(\Theta)\sin(\phi/2),
\qquad
I_{ge} = I_{eg} = i \sin(\Theta)\sin(\phi/2),
\end{equation}
and
\begin{gather}
\label{APPeq:OnePhotonFieldsTimeDomain}
\ket{\psi_{F}' (t)}= \int_{-\infty}^{t} \dd{t_1} \psi_1 (t_1) \ad(t_1) \ket{0_F},
\quad
\ket{\psi_{F}'' (t)}= \int_{t}^{+\infty} \dd{t_1} \psi_1 (t_1) \ad(t_1) \ket{0_F}, \nonumber
\\
\ket{\psi_{F,eg} (t)} = \int_{-\infty}^{t} \dd{t_1} \psi_1 (t_1) e^{-i \omega_q t_1} \ad(t_1) \ket{0_F},
\quad
\ket{\psi_{F,ge} (t)} = \int_{-\infty}^{t} \dd{t_1} \psi_1 (t_1) e^{+i \omega_q t_1} \ad(t_1) \ket{0_F}.
\end{gather}
We obtain Eq.~\eqref{eq:Onep_solution} by taking the limit $t\rightarrow +\infty$, and renaming $\ket{\psi_{F}' (t)}$ into $\ket{\psi_{F} (0)}$.

To compute the state of the qubit at time $t$ we exploit Eq.~\eqref{APPeq:ReducedStateSystemS} so that we have
\begin{equation}
\label{APPeq:QubitReducedStateOneZeroPhotons}
\rho_Q (t) = \prtqB{\int_{t}^{+\infty} \dd{t_1} \abs{\psi_1 (t_1)}^2}\rho_Q (0) + \int_{-\infty}^{t} \dd{t_1} \abs{\psi_1 (t_1)}^2 \tU_1 (t_1) \rho_Q (0) \tU_1^\dg (t_1),
\end{equation}
where $\rho_Q (0) = \dyad{\psi_Q (0)}$ is the initial state of the qubit.
In the basis $\prtg{\ket{e_\Theta},\ket{g_\Theta}}$ we have that:
\begin{multline}
\tU_1 (t_1) \rho_Q (0) \tU_1^\dg (t_1)
=\\
\mqty( \abs{b_e}^2\abs{I_{ee}}^2 + \abs{b_g}^2 \abs{I_{ge}}^2 - 2 \Re{b_g b_e^* I_{gg} I_{eg} e^{i t_1 \omega_q}} 
&
e^{i \omega_q t} I_{ee}I_{ge} \prt{\abs{b_g}^2 - \abs{b_e}^2} +
I_{ee}^2 b_e b_g^*+
b_e^* b_g \abs{I_{ge}}^2 e^{2 i \omega_q t}
\\
e^{-i \omega_q t} I_{gg}I_{ge} \prt{\abs{b_e}^2 - \abs{b_g}^2} +
I_{gg}^2 b_g b_e^*+
b_g^* b_e \abs{I_{ge}}^2 e^{-2 i \omega_q t} & 
\abs{b_g}^2\abs{I_{gg}}^2 + \abs{b_e}^2\abs{I_{eg}}^2 + 2 \Re{b_g b_e^* I_{gg} I_{eg} e^{i t_1 \omega_q}}
).
\end{multline}

\clearpage

\section{General formula of the spectrum\label{APPSec:SpectrumMultiPhoton}}

In this section we derive the formula of the light spectrum at any time. Notice that, in this Appendix, the initial time is taken to be $t=0$ instead of $t=-T$ in order to lighten the notation. The formulas reported in the main text are written considering the initial time to be $t=-T$.

We begin by rewriting the state of Eq.~\eqref{eq:SolutionGeneralDynamics} in the frequency domain by exploiting $a (t) = (1/\sqrt{2 \pi}) \intmp \dd{\omega} e^{-i \omega t} a (\omega)$ and $\ad(t) = (1/\sqrt{2 \pi}) \intmp \dd{\omega} e^{+i \omega t}\ad (\omega)$, obtaining (in the interaction picture with respect to field and system bare Hamiltonians)
\begin{equation}
\label{APPeq:FullDynamicsSolutionFrequencyDomain}
\ket{\psi (t)}
=
\prtq{\beta_0 + \sum_{n=1}^{\infty} \frac{\beta_n}{\sqrt{n!}}\intmp \dd{\omega^n} \Psi_n (\vec{\omega}_n,t) A^\dg_n (\vec{\omega}_n)}\ket{0_F}\ket{\psi_S (0)},
\end{equation}
where we defined
\begin{equation}
\Psi_n (\vec{\omega}_n,t) \equiv \frac{1}{(2 \pi)^{n/2}} \intmp \dd{t^n} \psi_n (\vec{t}_n)\tU_n (\vec{t}_n,t) e^{i \sum_j \omega_j t_j}.
\end{equation}
We now compute the spectrum $S(\omega,t) = \ev{\ad (\omega) a(\omega)}{\psi(t)}$ as follows:
\begin{equation}
S(\omega,t) = 
\sum_{n=1}^{\infty} \frac{\abs{\beta_n}^2}{n!} \intmp \dd{\omega}^n \dd{\gamma^n}
\ev{\Psi_n^\dg (\vec{\gamma}_n,t)\Psi_n (\vec{\omega}_n, t)}{\psi_S (0)}
\ev{A_n (\vec{\gamma}_n) \ad (\omega) a(\omega)  A^\dg_n (\vec{\omega}_n)}{0_F}
\end{equation}
The field average can be written exploiting Wick's theorem as follows:
\begin{equation}
\ev{A_n (\vec{\gamma}_n) \ad (\omega) a(\omega) A^\dg_n (\vec{\omega}_n)}{0_F}
=
\sum_{j,l=1}^n \delta (\omega_j - \omega)\delta(\gamma_l - \omega)
\prtq{\sum_{P(l'\neq l)} \prod_{j'\neq j} \delta (\omega_{j'} - \gamma_{P(l')})}.
\end{equation}
The number of terms in the above expression is $n^2 \times (n-1)! = n \times n!$. All the terms in the sum lead to the same integral because of the permutation invariance of the indices. Therefore, we can always reorder the variables in such a way that we can write
\begin{equation}
S(\omega,t) = 
\sum_{n=1}^{\infty} n \abs{\beta_n}^2 \intmp \dd{\gamma^{n-1}}
\ev{\Psi_n^\dg (\vec{\gamma}_{n-1},\omega,t)\Psi_n (\vec{\gamma}_{n-1},\omega, t)}{\psi_S (0)}.
\end{equation}
The final spectrum $S_{\rm out} (\omega)$ is obtained by taking the limit $t \rightarrow \infty$ in the above equation.

\subsection{Application to the qubit case}

In the case analyzed in the main text, we have a point-like qubit interacting with the field in the waveguide. We can write $\ket{\psi_S (0)} = b_g \ket{b_\Theta} + b_e \ket{b_\Theta}$, so that the final spectrum is always the sum of three parts:
\begin{equation}
    S_{\rm out} (\omega) = \abs{b_g}^2 S_{gg} (\omega) + \abs{b_e}^2 S_{ee} (\omega) + 2 \Re{b_g^* b_e S_{ge} (\omega)},
\end{equation}
where
\begin{equations}
S_{gg} (\omega) &= 
\sum_{n=1}^{\infty} n \abs{\beta_n}^2 \intmp \dd{\gamma^{n-1}}
\ev{\Psi_n^\dg (\vec{\gamma}_{n-1},\omega,t)\Psi_n (\vec{\gamma}_{n-1},\omega, t)}{g_\Theta},\\
S_{ee} (\omega) &= 
\sum_{n=1}^{\infty} n \abs{\beta_n}^2 \intmp \dd{\gamma^{n-1}}
\ev{\Psi_n^\dg (\vec{\gamma}_{n-1},\omega,t)\Psi_n (\vec{\gamma}_{n-1},\omega, t)}{e_\Theta},\\
S_{ge} (\omega) &= 
\sum_{n=1}^{\infty} n \abs{\beta_n}^2 \intmp \dd{\gamma^{n-1}}
\mel{g_\Theta}{\Psi_n^\dg (\vec{\gamma}_{n-1},\omega,t)\Psi_n (\vec{\gamma}_{n-1},\omega, t)}{e_\Theta}.
\end{equations}

Looking at Eq.~\eqref{eq:UnitaryOperator}, we can write (in the scattering limit)
\begin{equation}
\label{APPeq:UnitaryOperatorFactors}
\lim_{t\rightarrow +\infty}
e^{\frac{i t^{\uparrow}_j}{\hbar}H_0} e^{-\Theta_H (t-t_j^\uparrow)\frac{i \phi}{2}\sigma_z} e^{-\frac{i t^{\uparrow}_j}{\hbar}H_0}
=
I_{gg}\dyad{g_\Theta} + I_{ee}\dyad{e_\Theta} 
+ e^{-i \omega_q t^{\uparrow}_j} I_{ge}\dyad{g_\Theta}{e_\Theta} 
+ e^{+i \omega_q t^{\uparrow}_j} I_{eg}\dyad{e_\Theta}{g_\Theta},
\end{equation}
where $I_{\epsilon \varepsilon} \equiv \mel{\epsilon_\Theta}{e^{- i (\phi/2)\sigma_z}}{\varepsilon_\Theta}$ with $\epsilon,\varepsilon \in \lbrace e,g\rbrace$, their explicit values being
\begin{equation}
I_{gg} = I_{ee}^* = \cos(\phi/2)+ i \cos(\Theta)\sin(\phi/2),
\qquad
I_{ge} = I_{eg} = i \sin(\Theta)\sin(\phi/2).
\end{equation}
This implies that in a multi-photon pulse with a fixed number $n$ of photons, each photon can make the qubit change state or not. When it does, a phase is associated to the state, but notice that the $I_{\epsilon \varepsilon}$ are independent of $\omega$ or $t$. Therefore, we can write the following
\begin{equation}
\Psi_n (\vec{\omega}_n) \ket{g_\Theta} = 
\psi_{gg}^{(n)} (\vec{\omega}_n) \ket{g_\Theta} + 
\psi_{ge}^{(n)} (\vec{\omega}_n) \ket{e_\Theta},
\qquad
\Psi_n (\vec{\omega}_n) \ket{e_\Theta} = 
\psi_{ee}^{(n)} (\vec{\omega}_n)\ket{e_\Theta} + 
\psi_{eg}^{(n)} (\vec{\omega}_n) \ket{g_\Theta},
\end{equation}
where 
\begin{equations}
\psi_{gg}^{(n)} (\vec{\omega}_n) &= \sum_{\{\text{J.P.}\}} \mcI_{gg}^{(n)} (\text{J.P.}) \frac{1}{(2 \pi)^{n/2}} \intmp \dd{t^n} \psi_n (\vec{t}_n) \exp[- i \omega_q\sum_{\text{J.P.}} (-1)^{N[\text{J.P.}]} t^{\uparrow}_{\text{J.P.}}]e^{i \sum_j \omega_j t_j},\\
\psi_{ge}^{(n)} (\vec{\omega}_n) &= \sum_{\{\text{J.P.}\}} \mcI_{ge}^{(n)} (\text{J.P.}) \frac{1}{(2 \pi)^{n/2}} \intmp \dd{t^n} \psi_n (\vec{t}_n) \exp[- i \omega_q \sum_{\text{J.P.}} (-1)^{N[\text{J.P.}]} t^{\uparrow}_{\text{J.P.}}]e^{i \sum_j \omega_j t_j},\\
\psi_{ee}^{(n)} (\vec{\omega}_n) &= \sum_{\{\text{J.P.}\}} \mcI_{ee}^{(n)} (\text{J.P.}) \frac{1}{(2 \pi)^{n/2}} \intmp \dd{t^n} \psi_n (\vec{t}_n) \exp[+ i \omega_q \sum_{\text{J.P.}} (-1)^{N[\text{J.P.}]} t^{\uparrow}_{\text{J.P.}}]e^{i \sum_j \omega_j t_j},\\
\psi_{eg}^{(n)} (\vec{\omega}_n) &= \sum_{\{\text{J.P.}\}} \mcI_{eg}^{(n)} (\text{J.P.})\frac{1}{(2 \pi)^{n/2}} \intmp \dd{t^n} \psi_n (\vec{t}_n) \exp[+ i \omega_q \sum_{\text{J.P.}} (-1)^{N[\text{J.P.}]} t^{\uparrow}_{\text{J.P.}}]e^{i \sum_j \omega_j t_j}.
\end{equations}
In the above expressions, the $\mcI$ are numerical factors depending on $\Theta$ and $\phi$ while J.P stands for \enquote{Jump Points}, i.e., it denotes the precise sequence of jumps and non-jumps due to the interaction between photons and qubit. Using the notation of Sec.~\ref{Sec:MultiPhotonPulses} J.P. can stand, for example, for NJNJ, for a four-photon pulse. $N[\text{J.P.}]$ denotes instead the counting in the jump points. Then, in the four-photon pulse example of before, one has
\begin{equation}
\text{J.P.} = \text{NJNJ}
\implies
\exp[+ i \omega_q \sum_{\text{J.P.}} (-1)^{N[\text{J.P.}]} t^{\uparrow}_{\text{J.P.}}]
=
\exp[+ i \omega_q \prt{(-1)^{1} t^{\uparrow}_{2} + (-1)^{2} t^{\uparrow}_{4}}].
\end{equation}
Finally, $\{\text{J.P.}\}$ denotes the collection of possible jump points, so that, again using the notation introduced in Sec.~\ref{Sec:MultiPhotonPulses} for a pulse containing two photons, we get
\begin{multline}
	\psi_{gg}^{(2)} (\omega_1,\omega_2)
	=\\=
	\mcI_{gg}^{(2)} (GNN) \psi_{GNN} (\omega_1,\omega_2)
	+
	\mcI_{gg}^{(2)} (GNJ) \psi_{GNJ} (\omega_1,\omega_2)
	+
	\mcI_{gg}^{(2)} (GJN) \psi_{GJN} (\omega_1,\omega_2)
	+
	\mcI_{gg}^{(2)} (GJJ) \psi_{GJJ} (\omega_1,\omega_2).
\end{multline}

As the last example, we explicitly write the three-photon wavefunctions $\psi_{\text{GNJJ}}$ and $\psi_{\text{GNJJ}}$ (notice that they are not equal!):
\begin{equations}
\psi_{\text{GNJJ}} (\omega_1,\omega_2,\omega_3) &=  
\frac{1}{(2 \pi)^{3/2}} \intmp \dd{t_1}\dd{t_2}\dd{t_3} \psi_3 (t_1,t_2,t_3) \exp[- i \omega_q \prt{(-1)^1 t^{\uparrow}_{2}+(-1)^2 t^{\uparrow}_{3}}]\exp[i \prt{\omega_1 t_1 + \omega_2 t_2 + \omega_3 t_3}],\\
\psi_{\text{GJNJ}} (\omega_1,\omega_2,\omega_3) &=  
\frac{1}{(2 \pi)^{3/2}} \intmp \dd{t_1}\dd{t_2}\dd{t_3} \psi_3 (t_1,t_2,t_3) \exp[- i \omega_q \prt{(-1)^1 t^{\uparrow}_{1}+(-1)^2 t^{\uparrow}_{3}}]\exp[i \prt{\omega_1 t_1 + \omega_2 t_2 + \omega_3 t_3}].
\end{equations}
We recall that $t^{\uparrow}_{1}$ are the increasingly ordered time variables and are therefore functions of all time variables. In a more explicit notation, for example, $t^{\uparrow}_{1} (0.3,5,0) = 0$ and $t^{\uparrow}_{2} (0.3,5,0) = 0.3$.

The spectrum of a photonic wavefunction with exactly $n$ photons is equal to $n \rho_1 (\omega,\omega)$, where $\rho_1 (\omega,\omega')$ is the single-photon reduced density matrix in the frequency domain. In our case, to compute the spectrum of the wavefunctions given above, it is easier to first compute $\rho_1 (t,t')$, the single-photon reduced density matrix in the time-domain, and then Fourier transform it to get the spectrum.

\clearpage

\section{Dynamics and spectrum by means of the collision model approach\label{APPSec:CollisionModelherentState}}

In this section, we briefly describe how to derive the differential equations presented in sec.~\ref{sec:CoherentPulses} of the main text from a collision model approach. Notice that, in this Appendix, the initial time is taken to be $t=0$ instead of $t=-T$ in order to lighten the notation. The formulas reported in the main text are written considering the initial time to be $t=-T$.

First of all, we briefly show how to factorize a coherent state in discrete time-steps, starting by its definition~\cite{Book_Loudon2000}:
\begin{equation}
\ket{\alpha(t)} = \exp(\ad_\alpha - a_\alpha)\ket{0_F},
\qquad
\ad_\alpha 
\equiv \intmp \dd{t} \alpha(t) \ad (t)
\simeq \int_0^{2 t_0} \dd{t} \alpha (t) \ad (t)
\simeq \sum_{n=0}^{N} \alpha_n \ad_n,
\end{equation}
where we consider the pulse to be practically zero outside of $t \in [0,2t_0]$ and we discretized the time by steps of $\delta t = 2 t_0/N$. We also defined the quantity $\alpha_n \equiv \alpha(t_n)\sqrt{\delta t}$ and the operators $\ad_n = \ad (t_n) \sqrt{\delta t}$. Then, the discretization proceeds as follows~\cite{Ciccarello2017Collision,Maffei2022Closed}:
\begin{equation}
\ket{\alpha (t)} 
=
\exp(\ad_\alpha - a_\alpha)\ket{0_F}
\simeq
\exp(\sum_{n=0}^N \alpha_n \ad_n - \alpha_n^* a_n)\ket{0_F}
=
\prod_{n=0}^N \exp(\alpha_n \ad_n - \alpha_n^* a_n)\ket{0_F}
=
\bigotimes_{n=0}^{N} D(\alpha_n) \ket{0_n},
\end{equation}
where $D(\alpha_n)$ is the displacement operator acting on the ground state of the field in the $n$-th time cell, whose vacuum we denote by $\ket{0_n}$. When $N$ is high enough ($\delta t \rightarrow 0$), we can assume that we only need the ground and first excited level of each temporal mode. Hereafter in this Appendix, we will always work under this assumption.

The interaction between a time-cell and the qubit is given by
\begin{equation}
U_n = \exp[- i \frac{\omega_q}{2} \delta t \sigt - i \frac{\phi}{2} \sigma_z \hn_n] 
\end{equation}
where $\hn_n$ is the number operator for the $n$-th discretized temporal mode. Then, for the qubit evolution, after tracing out the field temporal mode, we get
\begin{equation}
\rho_{n+1} \simeq e^{-\abs{\alpha_n}^2}\prtq{K_0 \rho_n K_0^\dg + \abs{\alpha_n}^2 K_1 \rho_n K_1^\dg},
\qq{where}
K_0 \equiv e^{- i \frac{\omega_q}{2} \delta t \sigt},
\quad
K_1 \equiv \exp[- i \frac{\omega_q}{2} \delta t \sigt - i \frac{\phi}{2} \sigma_z] \simeq e^{- i \frac{\phi}{2}\sigma_z}.
\end{equation}
From the above equations we can now obtain differential equations by considering that
\begin{equation}
\label{APPeq:ExpansionsCollisionModel}
e^{-\abs{\alpha_n}^2} \simeq 1 - \abs{\alpha(t_n)}^2 \delta t,
\qquad
K_0 \simeq 1 - i \frac{\omega_q}{2}\sigt \delta t,
\qquad
K_1 \simeq e^{- i \frac{\phi}{2}\sigma_z} - i\frac{\omega_q}{2}\delta t \prtq{\cos(\Theta) \sigma_z e^{- i \frac{\phi}{2}\sigma_z} + \sin(\Theta) \frac{\sin(\phi/2)}{\phi} \sigma_x}.
\end{equation}
By inserting the above expansions and then dividing by $\delta t$ we get [also considering that $t = n \delta t$ and $\rho_n \rightarrow \rho(t)$]
\begin{equation}
\label{APPeq:QubitMasterEquationCoherentPulse}
\dot{\rho} (t) \equiv \lim_{\delta t \rightarrow 0} \frac{\rho_{n+1}-\rho_n}{\delta t} = - i \frac{\omega_q}{2}\comm{\sigt}{\rho(t)} + \abs{\alpha(t)}^2 \prtq{e^{- i \frac{\phi}{2}\sigma_z} \rho(t) e^{+i \frac{\phi}{2}\sigma_z} - \rho(t)}.
\end{equation}

To compute the phase quadrature $\Im\prtq{\ev{a(t)}}$ we have to compute the discretized operator $a_n$ on the coherent state at the end of its interaction with the qubit. Thus, we get
\begin{equation}
\ev{a(t_n)}
\simeq 
\frac{1}{\sqrt{\delta t}} \Tr \prtg{a_n U_n (\rho_n \otimes \dyad{\alpha_n}) U_n^\dg}
\simeq
e^{-\abs{\alpha_n}^2}\frac{\alpha_n}{\sqrt{\delta t}}\Tr{K_1 \rho_n K_0^\dg}.
\end{equation}
Again, from the above formula we can get a continuous version by inserting the expansions of Eq.~\eqref{APPeq:ExpansionsCollisionModel} and keeping only the zeroth-order terms we get:
\begin{equation}
\ev{a_{\rm out}(t)} = \alpha (t) \Tr{e^{- i \frac{\phi}{2}\sigma_z} \rho(t)}.
\end{equation}

Regarding the computation of the final spectrum, to obtain it we first compute the autocorrelation matrix $\ev{\ad(t_m) a(t_n)} \simeq (1/\delta t)\ev{\ad_m a_n}$ and then compute the spectrum from it. We get
\begin{multline}
\ev{\ad_m a_n}
=
\Tr \prtg{\ad_m a_n \prt{\prod_{j=0}^N U_j} \rho_0 \prt{\bigotimes_{j=0}^N \dyad{\alpha_j}} \prt{\prod_{j=0}^N U_j}^\dg}
=\\=
\Tr \prtg{\ad_m \prt{\prod_{j=n+1}^m U_j} \prtq{a_n U_n (\rho_n \otimes \dyad{\alpha_n}) U_n^\dg} \prt{\bigotimes_{j=n+1}^m \dyad{\alpha_j}} \prt{\prod_{j=n+1}^m U_j}^\dg}
=\\=
\alpha_n e^{-\abs{\alpha_n}^2}
\Tr \prtg{\ad_m U_m \mcE_{m,n} (K_1 \rho_n K_0^\dg) U_m^\dg}
=
\alpha_m^* \alpha_n e^{-\abs{\alpha_m}^2-\abs{\alpha_n}^2}
\Tr \prtg{K_0 \mcE_{m,n} (K_1 \rho_n K_0^\dg) K_1^\dg},
\end{multline}
where the super-operator $\mcE_{m,n} (\rho)$ represents the evolution due to the collisions between $n$ and $m$, i.e., 
\begin{equation}
\mcE_{m,n} (\rho) = \Tr_{F} \prtg{\prt{\prod_{j=n+1}^{m-1} U_j} \prt{\rho \bigotimes_{j=n+1}^{m-1} \dyad{\alpha_j}} \prt{\prod_{j=n+1}^{m-1} U_j}^\dg}.
\end{equation}
The above formula can be again greatly simplified by inserting the expansions of Eq.~\eqref{APPeq:ExpansionsCollisionModel} and keeping only the zeroth-order terms. We get:
\begin{equation}
\ev{\ad (s) a (t)} = \alpha^* (s) \alpha(t) \Tr \prtg{e^{+i \frac{\phi}{2}\sigma_z} \mcE_{t,s} \prtq{e^{- i \frac{\phi}{2}\sigma_z}\rho(t)}},
\end{equation}
where the super-operator $\mcE_{t,s}$ evolves its argument from $t$ to $s$ according to Eq.~\eqref{APPeq:QubitMasterEquationCoherentPulse}.

We can go further in our theoretical analysis by investigating the density matrix dynamics in the Bloch basis. We write $2\rho(t) = \Id + X(t)\sigma_x + Y(t)\sigma_y + Z(t) \sigma_z$ and we arrive at the following equation system:
\begin{equations}
\dot{X} &= -\abs{\alpha(t)}^2 \prtq{1 - \cos(\phi)}X -\prtq{\cos(\Theta) \omega_q + \abs{\alpha (t)}^2 \sin(\phi)} Y - \sin(\Theta) \omega_q Z,\\
\dot{Y} &= \prtq{\cos(\Theta) \omega_q + \abs{\alpha(t)}^2 \sin(\phi)}X -\abs{\alpha(t)}^2 \prtq{1 - \cos(\phi)} Y,\\
\dot{Z} &= \sin(\Theta) \omega_q Y.
\end{equations}
Moreover, we notice that
\begin{gather}
e^{- i \frac{\phi}{2}\sigma_z}\sigma_x = \cos(\phi/2)\sigma_x + \sin(\phi/2) \sigma_y,\qquad
e^{- i \frac{\phi}{2}\sigma_z}\sigma_y = -\sin(\phi/2)\sigma_x + \cos(\phi/2) \sigma_y,\\
e^{- i \frac{\phi}{2}\sigma_z}\sigma_z = \cos(\phi/2)\sigma_z -i\sin(\phi/2) \Id,\qquad
e^{- i \frac{\phi}{2}\sigma_z}\Id = -i\sin(\phi/2)\sigma_z + \cos(\phi/2)\Id.
\end{gather}
It follows that the formula for the quadrature can be written as
\begin{equation}
\ev{a_{\rm out}(t)} 
= \alpha (t) \Tr{e^{- i \frac{\phi}{2}\sigma_z} \rho(t)}
= \prtq{\cos(\phi/2) - i \sin(\phi/2)Z(t)}\alpha (t),
\end{equation}
which shows that only the imaginary part depends on the qubit's state.

\clearpage

\section{WAY theorem\label{APPsec:WAYTheorem}}

In their seminal works, Wigner~\cite{Busch2010translation}, Araki~\cite{Araki1960}, and Yanase~\cite{Yanase1961Optimal} showed that an observable that does not commute with an additive conserved quantity of the joint system (measured system plus quantum meter) cannot be measured projectively, projectively meaning with perfect accuracy and repeatability. This result goes under the name of the WAY (Wigner-Araki-Yanase) theorem and is at the origin of an active research field with a vast literature~\cite{Ozawa2002Conservation,Luo2003Wigner,Miyadera2006WAYtheorem,Loveridge2011Measurement,Navascues2014HowEnergy,Kuramochi2022wignerarakiyanase,Tajima2022Universal} dealing with the problem of setting fundamental constraints to the accuracy of quantum measurements. Within an autonomous meter-system interaction, the additive conserved quantity is the sum of the bare meter's and system's energies. The WAY theorem shows that, in this case, the measurement's quality is bounded by a function of the meter's energy variance~\cite{Ozawa2002Conservation,Loveridge2011Measurement}: it can be fully accurate and repeatable, i.e., projective, only when this energy variance is much higher than the system's bare energy. It follows that a macroscopic measurement apparatus can still achieve a projective measurement. The limitations given by the WAY theorem literature have practical consequences for general quantum technologies. Although macroscopic meters make these limitations negligible, this is not so when quantum meters (sometimes called ancillas) are used~\cite{Ozawa2002ConservativeQuantumComputing,Katsube2023limitations}. For convenience of the reader, we present here the main concepts related to the WAY theorem, adapted to our specific situation.

Let us consider a quantum measurement aiming to determine whether a qubit is in the $\ket{+}$ or $\ket{-}$ eigenstate of an observable $O_S$. The qubit (S) is coupled with its quantum meter (M) forming a closed bipartite system with Hamiltonian $H_{SM}=H_{S}+H_{M}+V$, with $H_{S(M)}$ being the qubit's (meter's) bare Hamiltonian and $V$ being their interaction. The system we consider is autonomous and in such a way that the interaction $V$ turns on and off autonomously during the dynamics. In other words, we consider a scattering type dynamics. Input and output states are then connected by the scattering operator $\mcS$, which is formally defined as $\mcS \equiv \lim_{t\rightarrow +\infty} U(+t,-t)$, where $U(t_2,t_1)$ is the unitary operator between times $t_1$ and $t_2$ in the interaction picture with respect to $H_0 = H_S + H_M$.  It is a general result of scattering theory that $\comm{H_0}{\mcS} = 0$~\cite{Book_Taylor2006Scattering}. In practice, the scattering operator can be defined as $\mcS = U(T,-T)$ because $V\sim 0$ for $t\leq -T$ and $t\geq T$.

At the initial time $t = -T$, the joint system is in a product state with $\ket{\psi_{M}(-T)}$ being the initial meter's state. An ideal pre-measurement dynamics leads to:
\begin{equations}
\label{eq:IdealPremeasurement}
U(T,-T) \ket{+}\ket{\psi_{M}(-T)} &= \ket{+}\ket{\psi_{M,++}},
\\
U(T,-T) \ket{-}\ket{\psi_{M}(-T)} &= \ket{-}\ket{\psi_{M,--}},
\end{equations}
with $\ip{\psi_{M,++}}{\psi_{M,--}}=0$. A classical measurement on the meter let us infer with perfect accuracy what was the qubit state at $t=-T$ and a subsequent measurement would give the same result. However, more in general, their interaction leads to the pre-measurement map:
\begin{equations}
\label{eq:general_map}
&U(T,-T) \ket{+}\ket{\psi_{M}(-T)}= c_{++}\ket{+}\ket{\psi_{M,++}}+c_{+-}\ket{-}\ket{\psi_{M,+-}};
\\
&U(T,-T) \ket{-}\ket{\psi_{M}(-T)}= c_{-+}\ket{+}\ket{\psi_{M,-+}}+c_{--}\ket{-}\ket{\psi_{M,--}}.
\end{equations}
We say that the pre-measurement is repeatable when $c_{+-}=c_{-+}=0$ as it leaves unaffected the eigenstates of $O_S$ so that a subsequent ideal measurement would get the correct result; it is accurate when $\ip{\psi_{F,++}}{\psi_{F,--}}=0$ as the states $\ket{\psi_{F,--}}$ and $\ket{\psi_{F,++}}$, i.e., the pointer states, can be perfectly distinguished by some classical measurement performed on the quantum meter later on~\cite{Fuchs1999Cryptographic} and thus determining with perfect accuracy what was the qubit's state at $t=-T$. 

Let us denote with $L_S$ a second qubit's operator, and with $L_M$ a meter's operator; the WAY theorem~\cite{Busch2010translation} states that perfect accuracy and repeatability cannot be simultaneously attained if (i) $\comm{L_S }{O_S} \neq 0$, and (ii) $\comm{L_S + L_M}{\mcS} =0$. This seminal result, corroborated by later and more quantitative results~\cite{Ozawa2002Conservation,Loveridge2011Measurement}, shows that accuracy and repeatability of the measurement generally increase when increasing the dispersion of the meter's operator $L_{M}$, until getting to the limit of the classical measurement apparatus where a nearly projective measurement can be achieved even if (i) and (ii) are verified. Indeed, since in our system $\comm{H_S + H_M}{\mcS} =0$, the above results apply upon the identification $L_S = H_S$ and $L_M = H_M$. 

\clearpage


%

\end{document}